\documentclass[reprint,showpacs,amsmath,amssymb,aip,jcp]{revtex4-1}
\usepackage{graphicx}
\usepackage{bm}
\usepackage{color} \definecolor{darkblue}{rgb}{0,0,0.6}
\usepackage[linkcolor=darkblue,citecolor=darkblue, colorlinks=true,breaklinks=true]{hyperref}

\begin{document}
\title{Direct Test of Supercooled Liquid Scaling Relations}

\author{T. Hecksher}\thanks{These authors contributed equally.}
\affiliation{Glass and Time, IMFUFA, Dept. of Science and Environment,
  Roskilde University, DK-4000 Roskilde, Denmark}

\author{D. H. Torchinsky}\thanks{These authors contributed equally.}
\affiliation{Department of Physics, MIT, Cambridge, MA 02139, USA}

\author{C. Klieber}\thanks{These authors contributed equally.}
\affiliation{Department of Chemistry, MIT, Cambridge, MA 02139, USA}

\author{J. A. Johnson}\thanks{These authors contributed equally.}
\affiliation{Department of Chemistry, MIT, Cambridge, MA 02139, USA}

\author{J. C. Dyre} \affiliation{DNRF Centre Glass and Time, IMFUFA,
  Dept. of Sciences, Roskilde University, DK-4000 Roskilde, Denmark}

\author{Keith A. Nelson} \affiliation{Department of Chemistry, MIT,
  Cambridge, MA 02139, USA}


\begin{abstract}
  Diverse material classes exhibit practically identical behavior when
  made viscous upon cooling toward the glass transition, suggesting a
  common theoretical basis. The first-principles scaling laws that
  have been proposed to describe the evolution with temperature have
  yet to be appropriately tested due to the extraordinary range of
  time scales involved. We used seven different measurement methods to
  determine the structural relaxation kinetics of a prototype
  molecular glass former over a temporal range of 13 decades and over
  a temperature range spanning liquid to glassy states. For the
  material studied, our results comprise a comprehensive validation of
  the two scaling relations that are central to the fundamental
  question of whether supercooled liquid dynamics can be described
  universally. The ultrabroadband mechanical measurements demonstrated
  have fundamental and practical applications in polymer science,
  geophysics, multifunctional materials, and other areas.
\end{abstract}
\maketitle

The extraordinary slowing down of viscous liquid dynamics upon cooling
toward the glassy state plays a key role in myriad contexts including
polymer processing, survival of living organisms in extreme cold,
amorphous metal synthesis, and many others. Glass-forming liquids
display a wide array of common features, despite quite different
chemistry ranging from high-temperature covalently bonded glass
formers to van der Waals liquids that typically form glasses below
room temperature \cite{Kauzmann1948, Angell1995, Debenedetti1996,
  Dyre2006, Roland2010, Berthier2011}.

\begin{figure}
  \begin{center}
    \includegraphics[width=7.2cm]{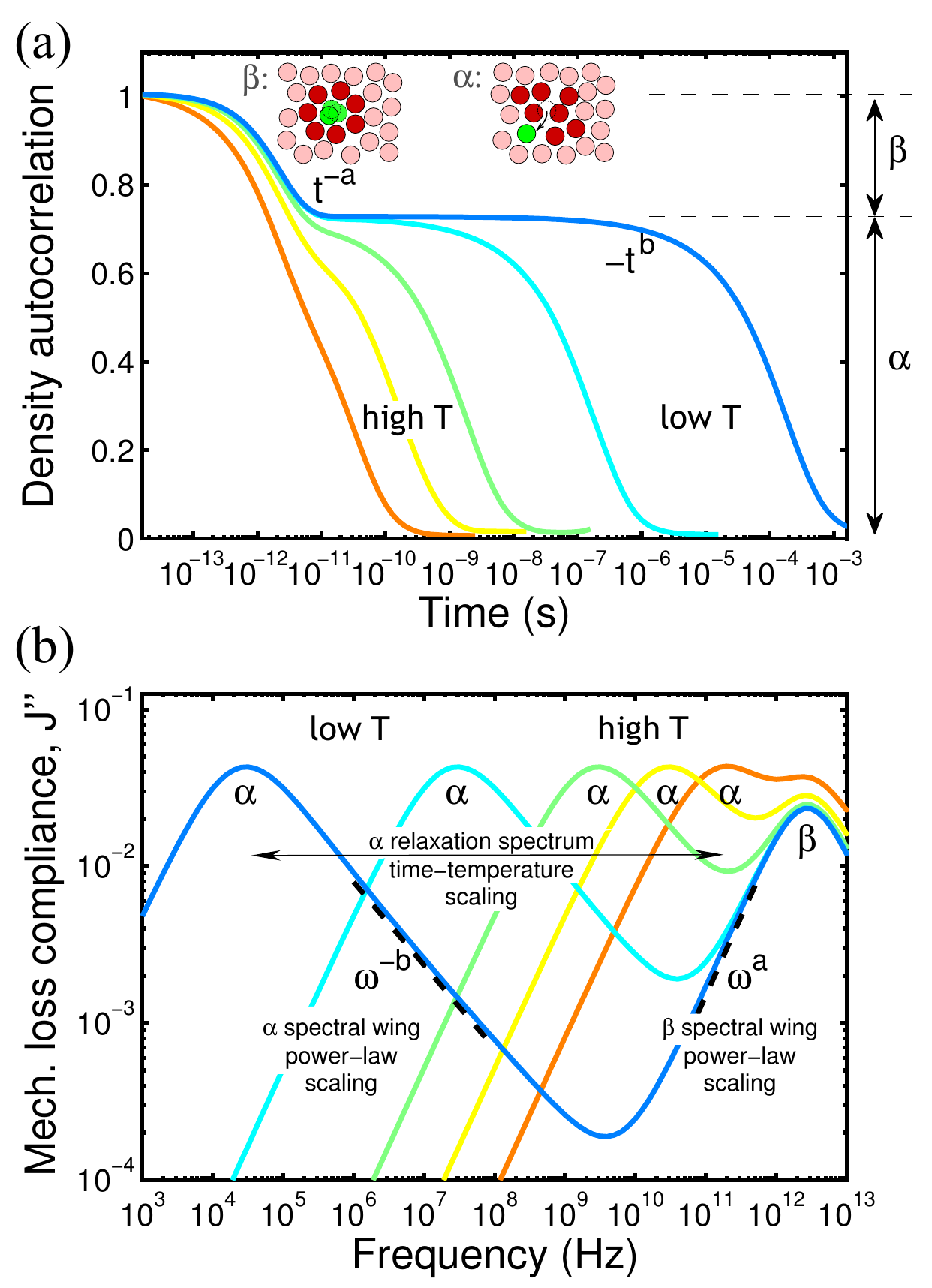}
  \end{center}
  \caption{\label{fig:correlators} Schematic depictions of primary
    (alpha) and fast (beta) relaxation processes in glass-forming
    liquids. Alpha and beta relaxation dynamics in (a) the time domain
    (density-density correlation function) and (b) the frequency
    domain (elastic compliance loss spectrum) at temperatures above
    the critical temperature $T_c$ predicted by mode-coupling theory
    (MCT). Time-temperature superposition is illustrated as the alpha
    relaxation spectral shape and width remain constant while the
    frequency range shifts by many decades as the temperature is
    varied. Dynamic critical exponents $a$ and $b$ describe the
    asymptotic power-law scaling of the relaxation kinetics predicted
    by MCT. At temperatures near $T_c$ there is a clear separation
    between alpha and beta relaxation time scales, with beta power-law
    relaxation of the density toward a time-independent ``plateau"
    value followed eventually by alpha power-law relaxation
    kinetics. In the frequency domain, the relaxation spectral wings
    show corresponding power-law frequency dependences around a
    minimum in the compliance loss spectrum. The drawings in (a) show
    a schematic representation of the beta relaxation among multiple
    potential energy minima (the different locations sampled by the
    central molecule colored green) within pre-existing intermolecular
    geometries or ``cages'' (formed by the red-colored molecules) and
    alpha relaxation involving larger-scale rearrangement of the
    intermolecular geometries to permit molecular diffusion and
    flow. }
\end{figure}

Viscoelastic relaxation behavior derives from two distinct and
sequential processes common to all glass forming liquids. The primary
or ``alpha'' structural relaxation dynamics, which dictate the time
scales for molecular diffusion and flow, are non-exponential in time,
typically extend over several decades of time scales at a single
temperature, and shift dramatically from picoseconds at high
temperatures and low viscosities to many seconds as the sample is
cooled and the glassy state is approached. This behavior gives rise to
broad loss peaks in elastic compliance spectra, covering an extended
frequency range at any temperature and shifting from gigahertz to
millihertz frequencies as the temperature is lowered (see
Fig.~\ref{fig:correlators}). In addition to the temperature-dependent
alpha relaxation dynamics for reorganization of intermolecular
geometries, fast local rearrangements of molecules within existing
``cage'' geometries, the so-called ``beta'' relaxation processes,
result in a higher-frequency feature in the loss spectrum. This is the
simplest scenario, which appears to obtain when structural relaxation
is slowed down through obstruction among neighboring molecules but not
through extensive networks as in hydrogen-bonded liquids or
entanglements as in polymers. In those more complicated cases,
additional processes between these two can be observed
\cite{Johari1970,Johari1982}. For the present study, we chose a van
der Waals molecular liquid in order to examine the universal alpha and
beta relaxation processes without complications from additional
dynamics.

A variety of empirical models \cite{Dyre2006a, Gibbs1958, Cohen1979,
  Kivelson1995, Jung2005, Tripathy2009}, as well as the
first-principles mode-coupling theory (MCT) \cite{Gotze1992,Das2004},
have been developed in an effort to rationalize the universal features
of supercooled liquid dynamics. In its idealized form, MCT predicts
for the local density dynamics a critical temperature $T_c$ at which
there is a transition from ergodic behavior above $T_c$ to nonergodic
behavior below, corresponding to arrest into a metastable glassy
state. The theory predicts a scaling law referred to as ``time
temperature superposition'' (TTS). TTS implies that the alpha
relaxation spectrum retains the same width and shape as the
temperature is changed, even as the frequency range varies widely.  A
wide range of practices in polymer processing, rheology, aging, and
other areas are based on TTS \cite{Ferry1980, Plazek1996, Olsen2001,
  Narayanaswamy1971, Moynihan1976}, but it has never been tested
directly for mechanical properties across most of the time scales
spanned. TTS is consistent with earlier heuristic descriptions of
supercooled liquids extending from high temperature all the way to the
glass transition temperature. Mode-coupling theory also predicts
distinct power-law scaling of alpha and beta relaxation processes at
temperatures above $T_c$, with the dynamic exponents connected through
arithmetic relations reminiscent of those associated with other
critical phenomena. Thus, the seemingly disparate relaxation
processes, separated by decades, are predicted to be intimately
related.

The density dynamics of supercooled liquids could be characterized
through measurement of the temperature-dependent elastic compliance
spectrum, but due to various experimental challenges this has never
been done over anywhere near the full frequency range of interest in a
single material. Ultra-broadband dielectric measurements
\cite{Lunkenheimer2000} and depolarized light scattering spectra
\cite{Petzold2010} clearly show the alpha and beta relaxation
features, but these techniques measure the dynamics of molecular
dipoles or polarizabilities whose coupling to density is complicated
and potentially temperature dependent. The key features of supercooled
liquid structural relaxation dynamics that might reveal universal
behavior at least within some classes of glass-forming materials
(e.g., organic van der Waals liquids) have been tested through light
scattering and neutron scattering spectra (see, e.g., \cite{Li1992,
  Li1992a, Gotze1999, Wuttke1994, Shen2000}), but not through direct
measurement of density dynamics over a sufficient frequency range.

\section{Experiments and Results}

We used seven complementary methods, based on six experimental setups,
to compile ultrabroadband longitudinal compliance spectra for the
glass-forming liquid tetramethyl tetraphenyl trisiloxane (sold
commercially as DC704). Four of the methods were photoacoustic
techniques \cite{Yan1987a, Silence1992, Thomsen1986, Choi2005,
  Supplementary} through which acoustic waves in the frequency range ~
1~MHz--100~GHz were generated and detected optically. The
corresponding acoustic wavelength range was $\sim 20$~nm--2~mm. Longer
acoustic wavelengths are comparable to sample sizes, so for lower
frequencies in the range $\sim1$~mHz--100~kHz, dynamical mechanical
analysis methods involving piezo-ceramics that shear or compress the
entire sample \cite{Hecksher2013, Christensen1994b, Christensen1995}
were used. The measurements spanned more than 13 orders of magnitude
in frequency with less than three decades of gaps, yielding structural
relaxation dynamics in the temperature range 200--320~K. Our earlier
measurements in the low and high-frequency ranges \cite{Hecksher2013,
  Klieber2013} lacked sufficient frequency coverage for comprehensive
testing of the scaling predictions, but the gap was bridged by the
newly developed Nanosecond Acoustic Interferometry (NAI) method. The
experimental methods and data analysis through which we determined the
compliance spectra and the elastic modulus (the inverse of the
compliance) are described in the online Supporting Information
\cite{Supplementary}.

\begin{figure}
  \begin{center}
    \includegraphics[width=8cm]{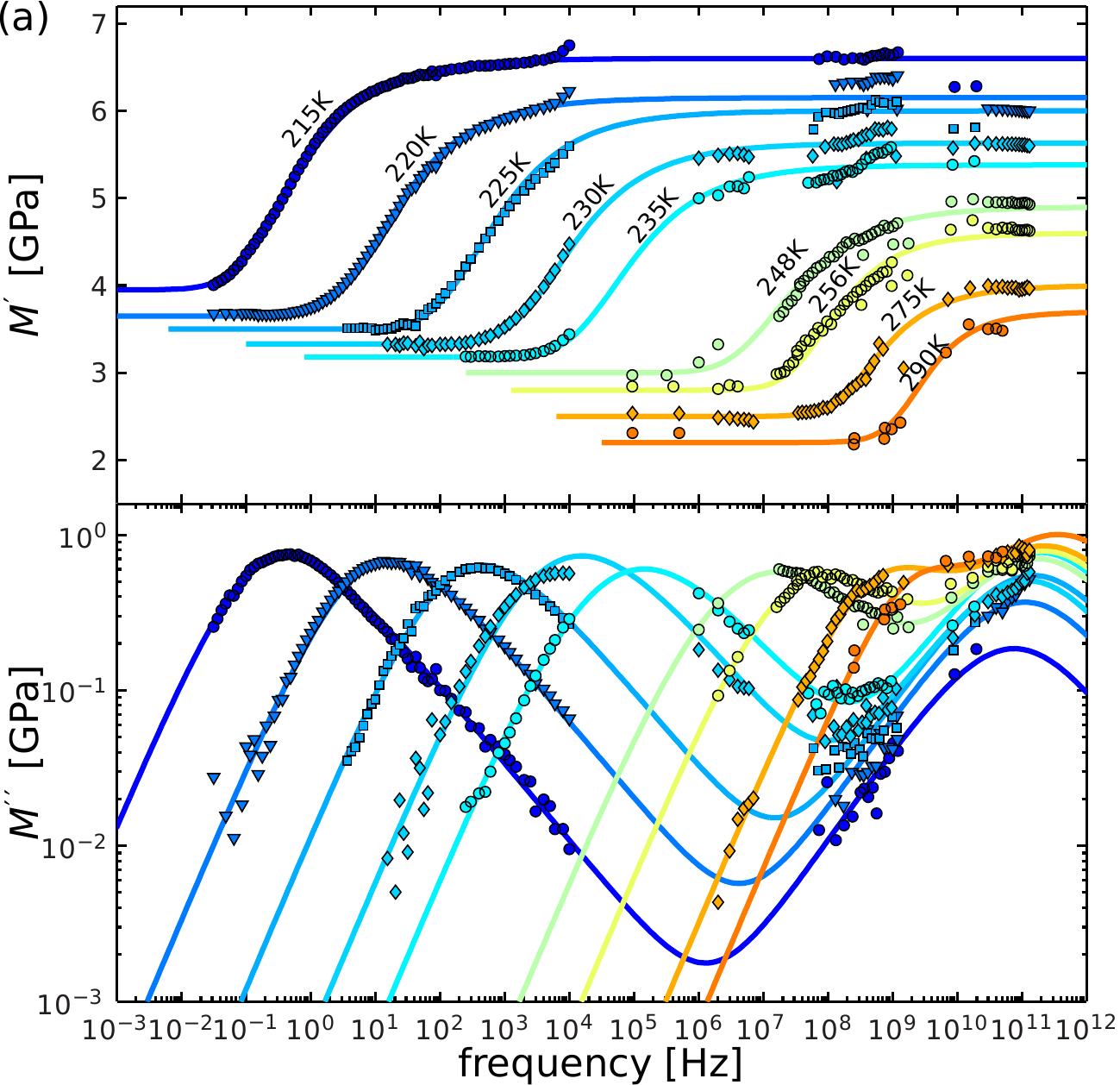}
    \includegraphics[width=8.5cm]{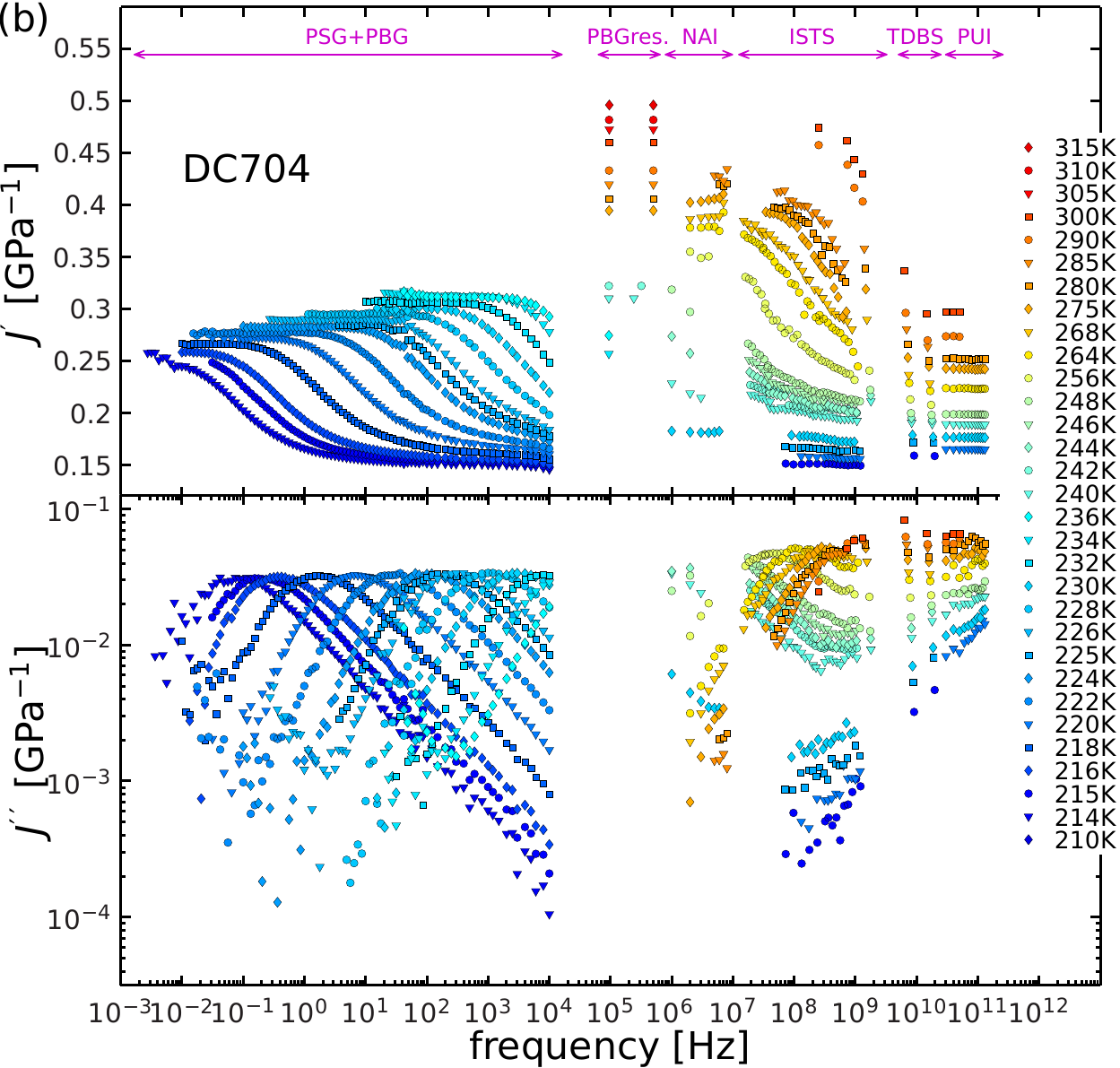}
  \end{center}
  \caption{\label{fig:modcom} (a) Longitudinal modulus spectra of
    DC704. Solid lines are guides to the eye. (b) Elastic compliance
    spectra of DC704. The techniques used in the different frequency
    ranges are indicated.}
\end{figure}

Figures~\ref{fig:modcom}(a) and \ref{fig:modcom}(b) present the
complex longitudinal modulus and compliance spectra of DC704 in the
frequency range of approximately $10^{-3}$–-$10^{11}$~Hz. The
imaginary parts of the spectra have gaps in the ranges of
approximately $10^4$–-$10^6$~Hz and 1–-10~GHz. In order to facilitate
visualization of the temperature-dependent trends, the modulus and
compliance spectra show only a subset of the data collected.

The primary temperature dependence, observed clearly in both real and
imaginary components of the modulus spectra, is the movement of the
alpha relaxation peak across about nine decades to lower frequencies
as the temperature is reduced toward the glass transition temperature
$T_g\approx 210$~K. At any given frequency $\omega_0$, the real part
of the modulus $M’(\omega_0,T)$ increases as the sample is cooled,
reflecting the stiffening of the liquid as it approaches the glassy
state. At any given temperature $T_0$, $M(\omega, T_0)$ plateaus at
frequencies above those of the alpha relaxation spectrum, since at
such high frequencies the liquid cannot undergo structural relaxation
during the acoustic oscillation period, resulting in a solid-like
response.

The compliance spectra show analogous behavior upon cooling: movement
of the alpha relaxation spectrum across many decades and a decrease in
the real part at low temperatures and high frequencies. The beta
relaxation feature is observed most clearly in the imaginary parts of
the modulus and compliance functions, which both rise at high
frequencies.

\begin{figure}
  \begin{center}
    \includegraphics[height=7cm]{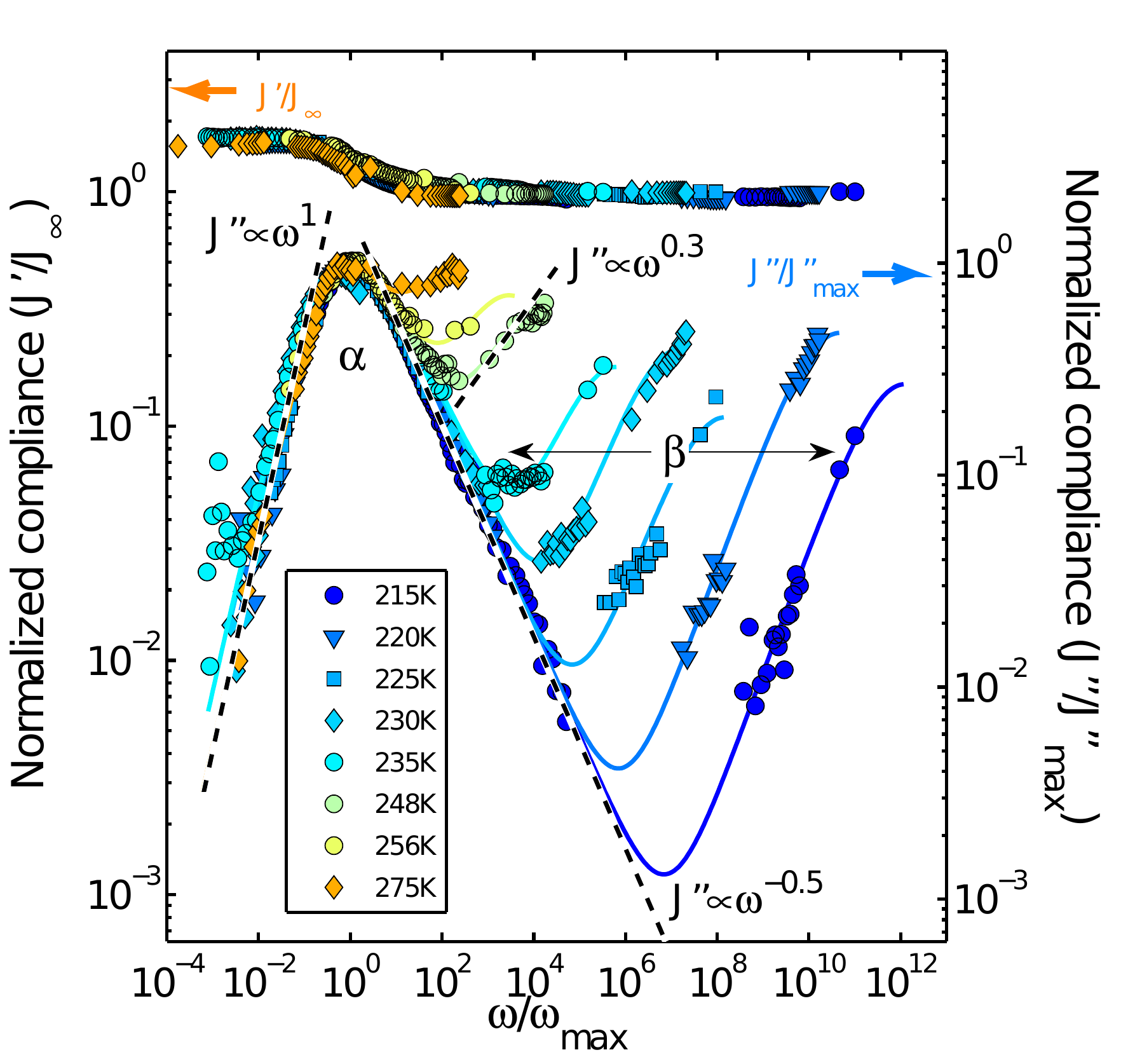}
  \end{center}
  \caption{\label{fig:tts}Master plot of the compliance demonstrating
    time-temperature superposition. The individual traces are scaled
    and shifted according to the fit values. The low-frequency side of
    the imaginary part follows a linear power-law, $J'' \propto
    \omega^1$ (dashed line), corresponding to exponential long-time
    kinetics. The full curves give analytic model fits based on an
    alpha relaxation spectrum that follows an $\omega^{-0.5}$
    high-frequency decay. The $\omega^{0.3}$ power-law also shown here
    corresponds to the value found to fit the beta relaxation spectral
    wing near the minimum at 248~K. (See Eq.~\ref{eq:min} and
    Fig.~\ref{fig:min})}.
\end{figure}

\section{Scaling Analysis}

Time-temperature superposition states that for the alpha relaxation
dynamics, the normalized density autocorrelation function can be
written as $\Phi(t,T) = F(t/\tau_{\alpha}(T))$, i.e., a single
temperature independent form $F(t/\tau_{\alpha}(T))$ describes the
relaxation dynamics at all temperatures, with the temperature
dependence contained only in the values of the characteristic
relaxation time $\tau_{\alpha}(T)$. In that case the alpha relaxation
spectrum $J''(\omega)$ displays analogous behavior. In order to test
TTS, we normalized the imaginary compliance spectra by the peak
heights and the real spectra by the high-frequency limiting value
$J_\infty$, and shifted both by amounts that made the peak frequencies
in $J''(\omega)$ coincide. The results are shown in
Fig.~\ref{fig:tts}. The alpha relaxation features superpose well
across the entire temperature regime studied. The imaginary parts have
the characteristic asymmetric shape found for most glass-forming
materials: the low-frequency sides of the alpha relaxation spectra
follow the Maxwell behavior, $J''(\omega) \propto \omega^1$, while the
high-frequency sides follow a power-law $J''(\omega) \propto
\omega^{-1/2}$. The shifting of the spectra to make the alpha peaks
overlap separates the beta relaxation features (which all appear at
the high-frequency range in Fig.~\ref{fig:modcom}) on the scaled
frequency axis. The scaled plots highlight the increase in separation
between alpha and beta relaxations upon cooling.

\begin{figure*}
  \begin{center}
    \includegraphics[height=7.cm]{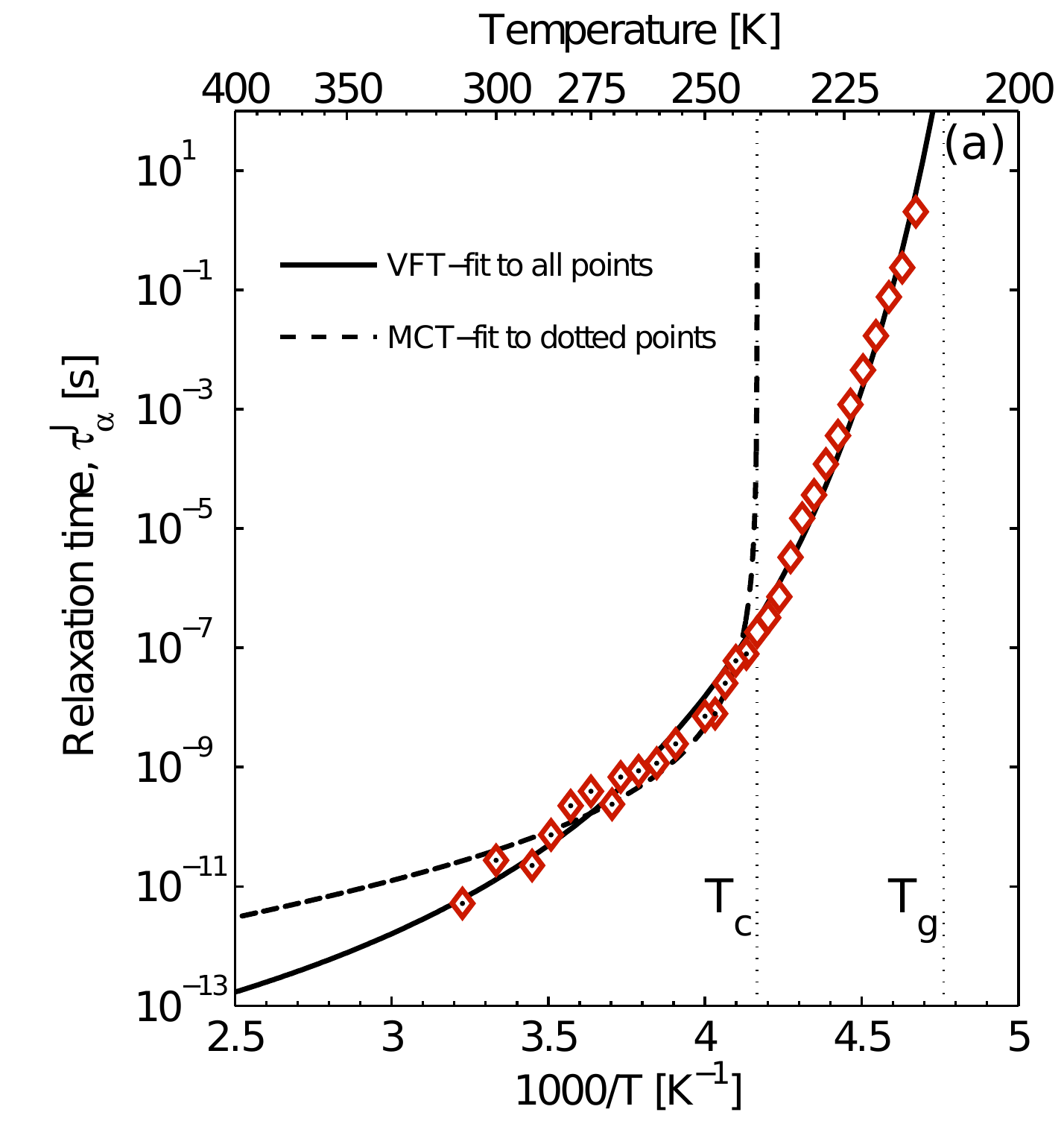}\hspace{.5cm}
    \includegraphics[height=7cm]{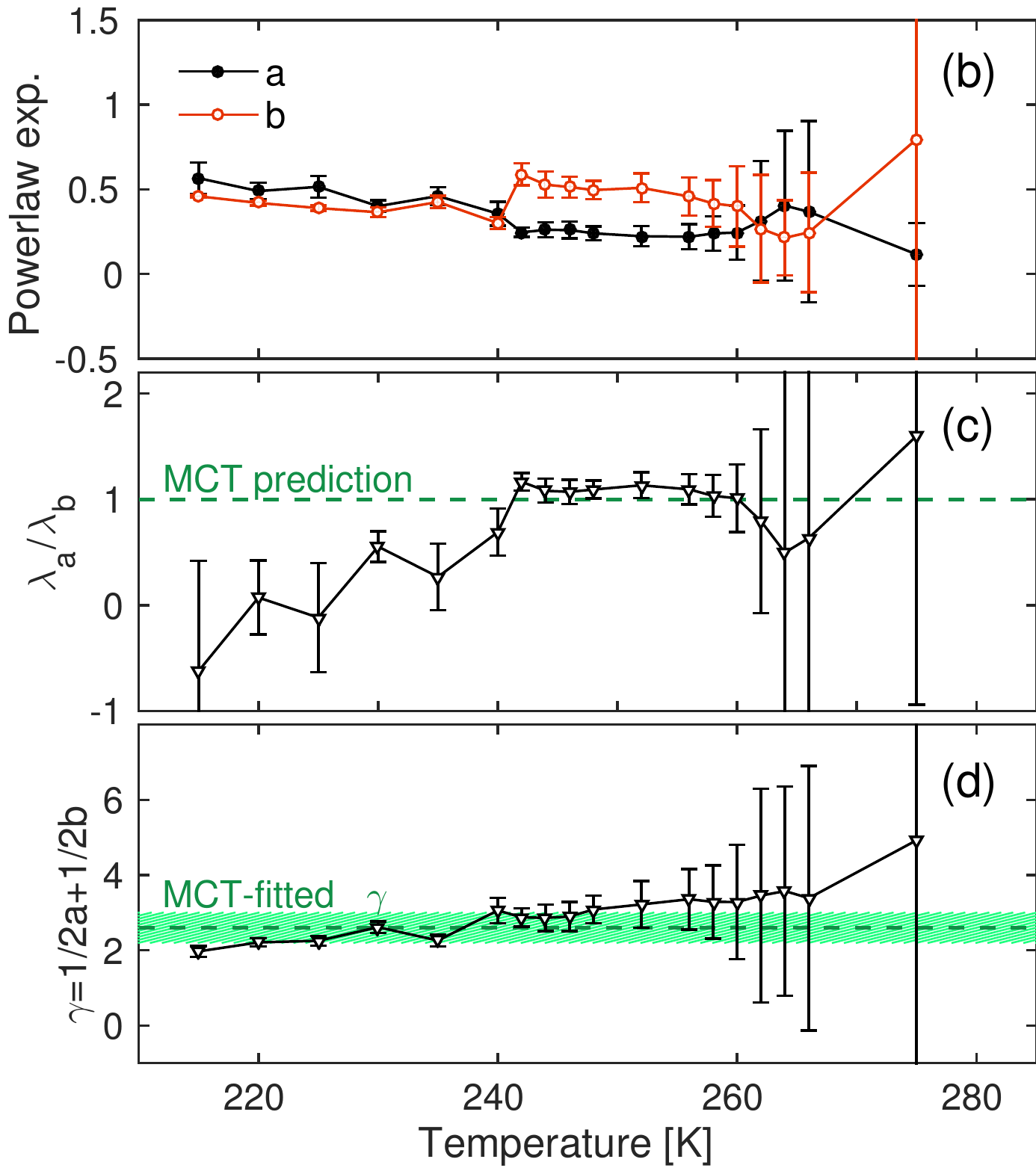}
  \end{center}
  \caption{\label{fig:mct} (a) The characteristic alpha relaxation
    time plotted as a function of inverse temperature where the
    Arrhenius equation gives a straight line. The glass transition
    temperature is at $T_g \cong$ 210~K. The VFT equation was fitted
    to all data points (full line) yielding $\log\tau_0 = -15.0\pm
    0.8$, $D=6\pm 1$ and $T_0 = 183 \pm 5$~K. The MCT critical
    temperature $T_c = 240$~K was identified from (b) as the temperature
    below which MCT predictions clearly break down; the MCT power-law
    fit to the high-temperature points $(T > T_c)$ (marked by central
    black dots) yielded the following fit parameters:
    $\log(\tau_x/\text{s}) = -12.0\pm 0.5$, $\gamma = 2.7\pm 0.4$. (b)
    Values of powerlaw exponents $a$ and $b$ over a wide temperature
    range, determined as fitting parameters to Eq. (3). (c) Test of
    the MCT prediction in Eq.~(\ref{eq:lambda}) that $\lambda_a =
    \lambda_b$. The test is useful over the 240--248~K temperature
    range. (d) The calculated exponent $\gamma$ as a function of
    temperature. The green dotted line shows the value of obtained
    from fitting Eq.~(\ref{eq:tau}) to the relaxation times (in (d))
    and the green area gives the estimated uncertainty on this value.}
\end{figure*}

Figure~\ref{fig:mct}(a) shows the values of the characteristic
relaxation time $\tau_\alpha$ of the alpha process as a function of
temperature determined from fits to the stretched exponential function
$\exp(-(t/\tau_{\alpha})^n)$ with a temperature-independent stretching
exponent $n$ and strongly temperature dependent $\tau_\alpha$
values. The stretching exponent was fixed at $n = 0.5$, which at high
frequencies corresponds to an $\omega^{-1/2}$ decay of the imaginary
part of the compliance \cite{Barlow1967, Dyre2006, Nielsen2009}. The
observed non-Arrhenius temperature dependence is typical of
glass-forming liquids and is associated generally with a complex
energy landscape rather than a single activation energy for all
relaxation processes \cite{Angell1995, Debenedetti1996, Dyre2006,
  Roland2010, Berthier2011, Bohmer1993}. The relaxation time data were
fit to the MCT power-law prediction\cite{Gotze1992}
\begin{equation}\label{eq:tau}
\tau_{\alpha} = \tau_x \left[T_c/(T-T_c)\right]^{\gamma}
\end{equation}
with $\tau_x$ and $\gamma$ as free parameters. This prediction is for
temperatures above $T_c$ (taken as 240~K as described below), although
as suggested heuristically we find that TTS applies also below
240~K. The full set of $\tau_{\alpha}(T)$ values was fitted to the
commonly used empirical Vogel-Fulcher-Tammann (VFT) function
\cite{Vogel1921,Tammann1925} $\tau_{\alpha}(T) = \tau_0 \exp
\left[DT_0/(T - T_0)\right]$.

The MCT interpolation formula \cite{Cummins1993}
\begin{equation}\label{eq:min}
  J''(\omega)=\frac{J''_{min}}{a+b}
  \left[a\left(\frac{\omega}{\omega_{min}}\right)^{-b} +
    b\left(\frac{\omega}{\omega_{min}}\right)^a\right] 
\end{equation}
connects the two relaxation features around the minimum between. This
expression was fitted to the measured spectra in the region around the
minimum. The fitted values for $a$ and $b$ are shown in
Fig.~\ref{fig:mct}(b). The uncertainty of the fits increases
dramatically as the temperature increases due to the merging of the
two processes making a distinction of the separate processes
difficult. The MCT predicts a relation between the two exponents
according to which
\begin{equation}\label{eq:lambda}
  \lambda_a = \frac{\Gamma(1-a)^2}{\Gamma(1-2a)}=\lambda_b =
  \frac{\Gamma(1+b)^2}{\Gamma(1+2b)}
\end{equation}
where $\Gamma(z)$ is the gamma-function defined as $\Gamma(z)
=\int_0^{\infty}x^{z-1}e^{-x}dx$. The relation is tested in
Fig.~\ref{fig:mct}(c), which shows that the data are consistent with
this prediction at temperatures above 240~K, but clearly breaks down
below. We take this to define the MCT critical temperature. The value
of the alpha relaxation time at $T_c$, $\tau_{\alpha}$(240~K)~$\approx
0.1~\mu$s, is typical for $T_c$, below which idealized MCT breaks down
\cite{Das2004} due to the onset of thermally assisted hopping
\cite{Goldstein1969,Debenedetti2001,Schroder2000} through which alpha
relaxation continues to occur.

Figure \ref{fig:mct}(d) compares the alpha relaxation critical
exponent $\gamma$ determined in Fig.~\ref{fig:mct}(a) based on
$T_c=240$~K to the MCT prediction that connects the power-law
exponents to $\gamma$ through the relation\cite{Gotze1992}
\begin{equation}\label{eq:gamma}
  \gamma =  1/(2a) + 1/(2b)\,.
\end{equation}

The above analysis indicates that MCT is valid at temperatures above
240~K for the studied liquid. For the fits to be useful as a test of
the MCT predictions, there must be sufficient separation between the
alpha and beta relaxation spectra as $T \rightarrow T_c$ from above
that the $a$ and $b$ exponents can be associated distinctly with the
two features\cite{Cummins1999,Gotze1999}. This was the case for our
results below 248~K and thus MCT can be meaningfully tested in the
temperature range 240--248~K.

\begin{figure}
  \begin{center}
    \includegraphics[width=8.5cm]{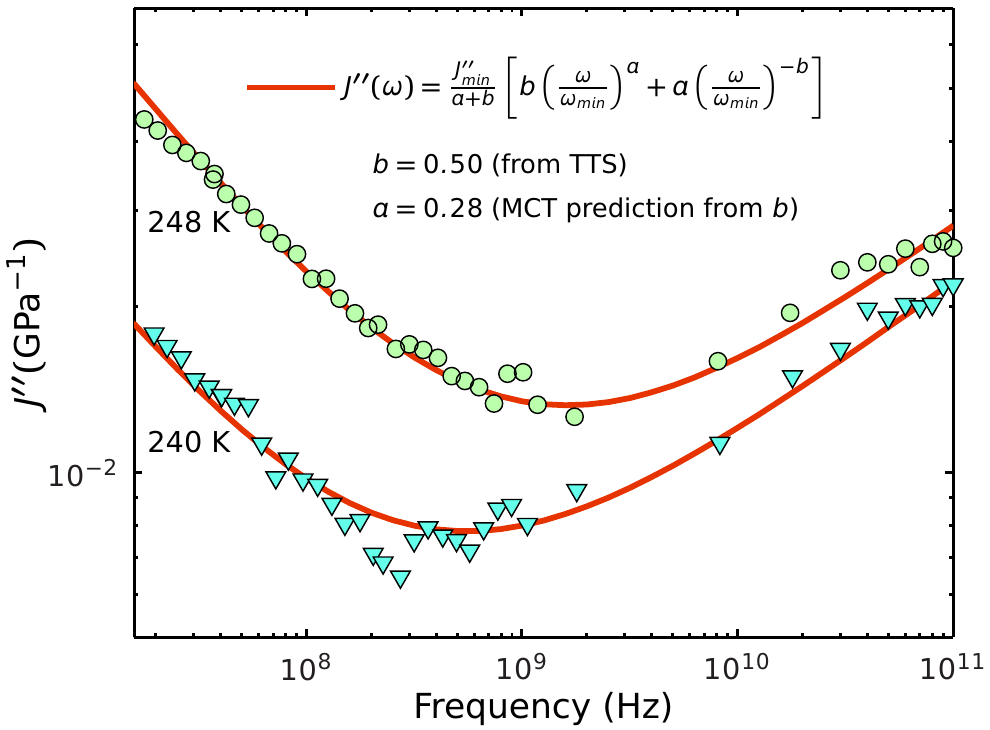}
  \end{center}
  \caption{\label{fig:min}Data around the minimum of the imaginary
    part of the compliance function fitted to the MCT power-law
    relation, Eq.~\ref{eq:min}, without adjustable shape parameters
    (critical exponents $a$ and $b$) at sample temperatures of 240~K
    and 248~K. The exponent $b = 0.50$ was determined from the
    low-temperature data using time-temperature superposition;
    Eq.~\ref{eq:lambda} then yielded $a = 0.28$. }
\end{figure}

Figure~\ref{fig:min} shows the region around the minima between the
alpha and beta features of the compliance loss spectra $J''(\omega)$
at temperatures 240~K and 248~K.  As a final test Eq.~(\ref{eq:min})
was compared to the data with fixed exponent values. Only the minimum
frequency $\omega_\text{min}$ and the minimum compliance loss
$J''_\text{min}$ positions of each curve were adjusted to fit the
data.  The exponent value $b = 0.50$ was determined from the
high-frequency wing of the alpha relaxation spectra evident in the TTS
plot (Fig.~\ref{fig:tts}). From this the predicted value of the
exponent $a$ could be calculated using Eq.~(\ref{eq:lambda}). This
yielded the value $a = 0.28$, which should describe the low-frequency
wing of the beta relaxation spectra. The resulting prediction
completely determines the shape of the curves which go through the
data well within the experimental scatter.

\section{Conclusions}

Our measurements provide the first ultra-broadband mechanical
relaxation spectra of a glass-forming liquid reaching the high
frequencies of the beta relaxation spectrum and extending to the low
frequencies of the alpha relaxation spectrum even as $T_g$ is
approached. The results give direct experimental support for
time-temperature superposition of the alpha relaxation spectrum above
and below the MCT critical temperature (240~K), the validity of which
is often assumed in empirical studies of glass-forming liquids and in
modeling used for practical applications. Our results also permitted
calculation of the dynamic critical exponents of MCT above $T_c$,
yielding results consistent with predictions that relate the alpha and
beta relaxation dynamics. In continuing work, we are filling in the
frequency gaps of just under three decades within the more than 13
decades that the present measurements cover. We are also extending the
measurements to higher frequencies in order to improve the reliability
of our tests. We also have measured shear relaxation dynamics across a
wide frequency range \cite{Klieber2013}, and filling in the frequency
gaps for these measurements will permit comparison between
longitudinal and shear dynamics that may show distinct
temperature-dependent behavior \cite{Torchinsky2012}.

DC704 is the first sample for which mechanical relaxation dynamics
have been measured across the frequency range we have
explored. Additional glass-forming liquids must be examined in order
to assess the generality of TTS and other MCT predictions. Our results
demonstrate that access is now available to the extraordinarily wide
frequency range needed for such comprehensive tests of supercooled
liquids and a wide range of partially disordered materials including
relaxor ferroelectrics, block copolymers, and many others. 

\section{Materials and methods}

\subsection{Overview}

The seven different measurement methods and the frequency ranges that
they cover are summarized in Fig.~\ref{fig:techniques_survey};
detailed descriptions of the techniques and data collected from them
are discussed in the Supporting Information \cite{Supplementary}. The
techniques include three low-frequency methods involving
piezo-ceramics that shear or compress the entire sample
quasi-statically \cite{Hecksher2013, Christensen1994b,
  Christensen1995, Hecksher_thesis} and four higher-frequency methods
utilizing short laser pulses to excite and subsequently detect
acoustic waves in an irradiated region \cite{Johnson_thesis, Yan1987b,
  Silence1992, Torchinsky_thesis, Thomsen1986, Klieber_thesis,
  Choi2005}.

The two lowest-frequency methods determine the complex
frequency-dependent bulk modulus $K(\omega)$ and shear modulus
$G(\omega)$ directly, where $\omega$ is the angular frequency. The
longitudinal modulus is then given by $M(\omega) = M'(\omega) +
iM''(\omega) = K(\omega) + (4/3)G(\omega)$. The four methods covering
MHz-GHz frequencies determine acoustic parameters. The complex
longitudinal modulus is given from the acoustic data as $M(\omega, T)
= \rho(T)(c_{L}(\omega, T))^2$ where $\rho(T)$ is the
temperature-dependent density and $c_{L}(\omega, T)$ the complex
frequency-dependent longitudinal sound velocity.

In order to determine the modulus as a function of temperature from
the sound velocity and damping rate, the thermal contraction of the
sample must be accounted for. Using literature data of the thermal
expansion coefficient $\alpha = 7.2\times
10^{-4}[\rm{K}^{-1}]$\cite{Orcutt1973,Poulter1979} and assuming this
quantity is temperature independent, the following expression for the
temperature dependence of the density can be derived
\cite{Klieber2013}.
\begin{equation}
  \begin{split}
    \rho(T) & = \frac{\rho(T_{ref})}{1+\alpha_p(T-T_{ref})} \\
    & = \frac{1.07~[{\rm kg/m^3}]}{1 + 7.2 \times 10^{-4}~[
      \rm{K}^{-1}]\left(T - 298[{\rm K}] \right)} \,
  \end{split}
\end{equation}
Once the modulus is obtained, the complex longitudinal compliance is
given by $J(\omega) = 1/M(\omega) = J'(\omega) - iJ''(\omega)$.

MCT relations are predictions for the mechanical susceptibility
$\chi(\omega)$ to which $J(\omega)$ is proportional. Strictly
speaking, $\chi(\omega)$ is related to the density autocorrelation
function and therefore to the bulk compliance, not the longitudinal
compliance. However, the difference between $K(\omega)$ and
$M(\omega)$ (and thus also the bulk and longitudinal compliances) is
negligible because the shear modulus is considerably smaller than the
longitudinal modulus, and their frequency-dependent dynamics are very
similar at both low \cite{Hecksher2013,Christensen1994,
  Gundermann2014} and high \cite{Torchinsky2012} frequencies.

\begin{figure}
  \centering
  \includegraphics[width=7.9cm]{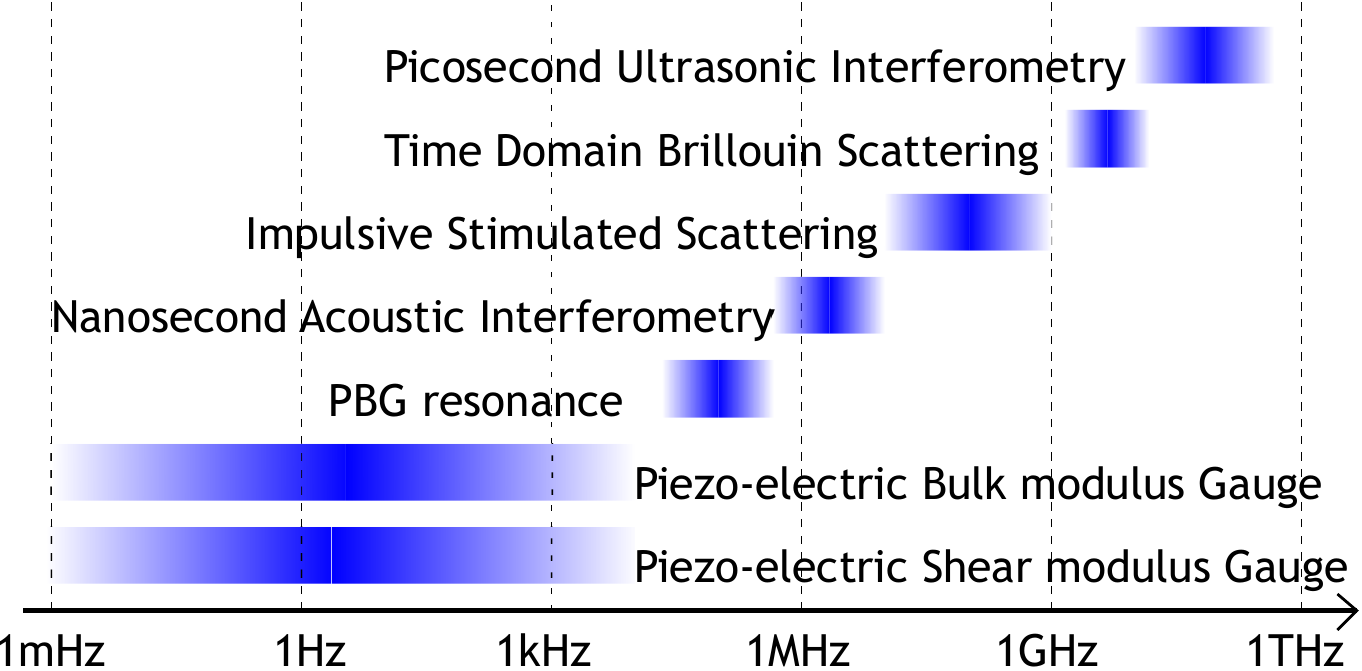}
  \caption{\label{fig:techniques_survey}Survey of mechanical
    spectroscopic techniques. At low frequencies (from mHz up to
    $\sim10$ kHz) the piezo-electric Shear modulus Gauge (PSG) and the
    piezo-electric bulk modulus Gauge (PBG) were used to measure the
    shear and bulk moduli. Analysis of the overtones in the PBG
    provided access to some data points in the $100-500$ kHz
    region. In the low MHz range, nanosecond acoustic interferometry
    (NAI) was used to probe the frequency dependence of the
    longitudinal sound speed and attenuation rates directly. In the
    higher MHz range, impulsive stimulated scattering (ISS) was used
    to measure longitudinal sound velocities and damping at specified
    acoustic wavevectors. Finally, time-domain Brillouin light
    scattering (TDBS) and picosecond ultrasonic interferometry (PUI)
    were used to measure longitudinal acoustic speeds and attenuation
    in the MHz-GHz frequency ranges shown.}
\end{figure}

\subsection{Sample Preparation}

For the PSG, PBG, and ISS measurements, DC704 was obtained from Sigma
Aldrich and used without purification. For the ISS measurements, the
liquid was transferred into the cell through a 0.22~$\mu$m millipore
filter into a fused quartz cuvette. During the ISS measurements, the
sample was observed to become slightly opaque at the coldest
temperatures. This is likely due to phase separation of dissolved
impurities from the base liquid and disappeared when the liquid was
reheated. The problem was overcome in subsequent experiments by mixing
the sample with anhydrous ${\rm MgSO_4}$, combined with heating under
vacuum before filtration. This approach was adopted for the NAI, TDBS,
and PUI techniques. Comparison of data both with and without treatment
by the drying agent showed no difference in the acoustic
parameters. The DC704 samples never crystallized during the course of
our experiments.

The transducer techniques utilize home-built closed-cycle cryostats
capable of keeping the temperature constant within a $\sim5$~mK
\cite{Igarashi2008a}, while the NAI and ISS measurements were
performed in a commercially available cold-finger cryostat. In these
measurements, temperature sensing was provided by a factory calibrated
platinum resistors immersed in the liquid a few millimeters away from
the optical beams. The TDBS and PUI techniques were performed in a
commercial sample-in-vapor cryostat and the temperature was monitored
at a position a few millimeters away from the sample.

We did not carry out any calibration of temperatures between the
cryostats of the different labs. However, the low-frequency part of
the longitudinal spectrum was obtained as the sum of two individual
measurements, which were carried out in the same experimental set-up
(same cryostat, same electronics). We estimate that the uncertainty on
the absolute temperature is less than the overall noise in the
high-frequency methods.

\begin{acknowledgments} 
  The work at MIT was supported in part by National Science Foundation
  Grant No. CHE-1111557 and Department of Energy Grant
  No. DE-FG02-00ER15087. The work at Roskilde University was sponsored
  by the DNRF Grant no 61.
\end{acknowledgments}

\pagebreak
\widetext
\begin{center}
\textbf{\large Supplementary Material: Direct Test of Supercooled Liquid Scaling Relations.}
\end{center}
\setcounter{equation}{0}
\setcounter{figure}{0}
\setcounter{table}{0}
\setcounter{page}{1}
\setcounter{section}{0}
\makeatletter
\renewcommand{\theequation}{S\arabic{equation}}
\renewcommand{\thefigure}{S\arabic{figure}}
\renewcommand{\bibnumfmt}[1]{[S#1]}
\renewcommand{\citenumfont}[1]{S#1}







\maketitle
\setcounter{equation}{0}
\setcounter{figure}{0}
\setcounter{table}{0}
\setcounter{page}{1}
\makeatletter
\renewcommand{\theequation}{S\arabic{equation}}
\renewcommand{\thefigure}{S\arabic{figure}}
\renewcommand{\bibnumfmt}[1]{[S#1]}
\renewcommand{\citenumfont}[1]{S#1}

\section{Experimental Methods}

Overview over the seven different measurement methods and the
frequency ranges that they cover are summarized in
Fig.~\ref{fig:techniques_survey}; detailed descriptions of the
techniques and data collected from them are discussed in the
subsequent sections below. The techniques include three low-frequency
methods involving piezo-ceramics that shear or compress the entire
sample quasi-statically\cite{sHecksher2013, sChristensen1994b,
  sChristensen1995, sHecksher_thesis} and four higher-frequency methods
utilizing short laser pulses to excite and subsequently detect
acoustic waves in an irradiated region\cite{sJohnson_thesis, sYan1987b,
  sSilence1992, sTorchinsky_thesis, sThomsen1986, sKlieber_thesis,
  sChoi2005}.

\begin{figure}[h!]
  \centering
  \includegraphics[width=7.9cm]{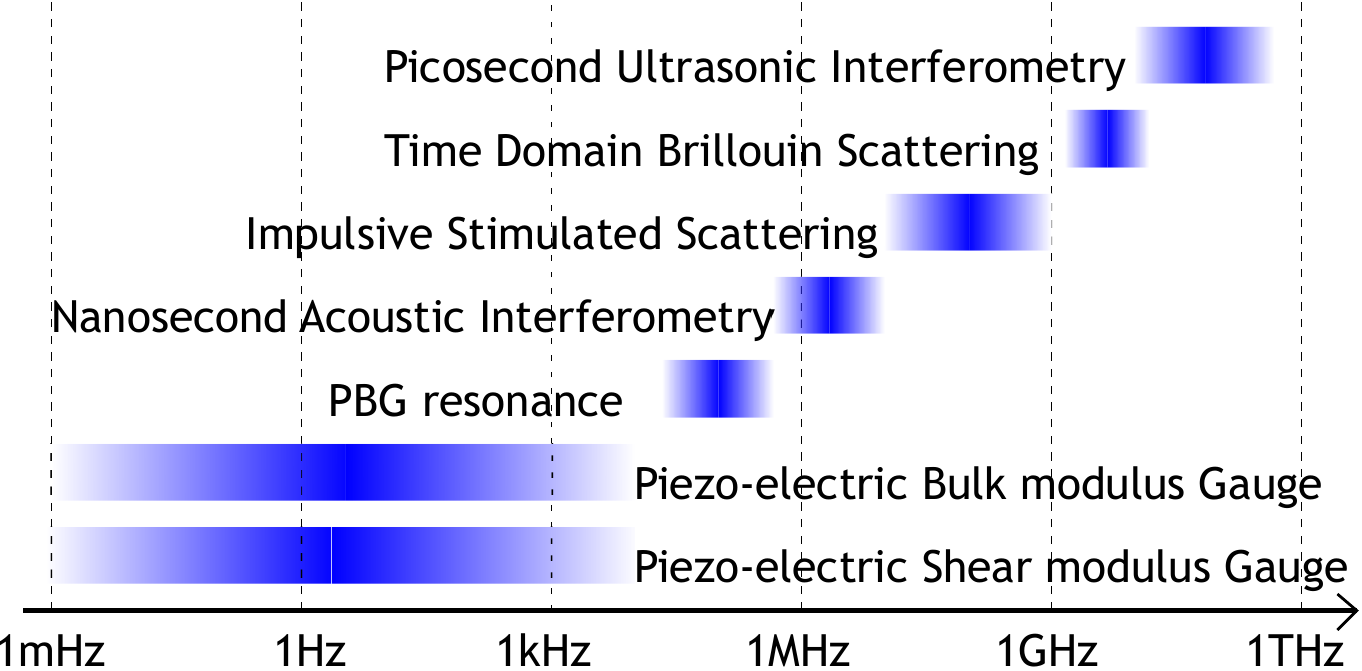}
  \caption{\label{fig:techniques_survey}Survey of mechanical
    spectroscopic techniques. At low frequencies (from mHz up to
    $\sim10$ kHz) the piezo-electric Shear modulus Gauge (PSG) and the
    piezo-electric bulk modulus Gauge (PBG) were used to measure the
    shear and bulk moduli. Analysis of the overtones in the PBG
    provided access to some data points in the $100-500$ kHz
    region. In the low MHz range, nanosecond acoustic interferometry
    (NAI) was used to probe the frequency dependence of the
    longitudinal sound speed and attenuation rates directly. In the
    higher MHz range, impulsive stimulated scattering (ISS) was used
    to measure longitudinal sound velocities and damping at specified
    acoustic wavevectors. Finally, time-domain Brillouin light
    scattering (TDBS) and picosecond ultrasonic interferometry (PUI)
    were used to measure longitudinal acoustic speeds and attenuation
    in the MHz-GHz frequency ranges shown.}
\end{figure}

\subsection{Low-frequency methods}\label{sec:PG}

The low-frequency methods measure mechanical moduli directly. These
techniques do not measure the longitudinal modulus, but the bulk and
shear moduli. In the isotropic case (e.g., in a liquid) there are only
two unique mechanical moduli, and the longitudinal modulus $M$ is
given in terms of the bulk ($K$) and shear ($G$) moduli as:
$M=K+4/3G$.

\begin{figure}[htpb!]
  \begin{center}
    \includegraphics[width=8cm]{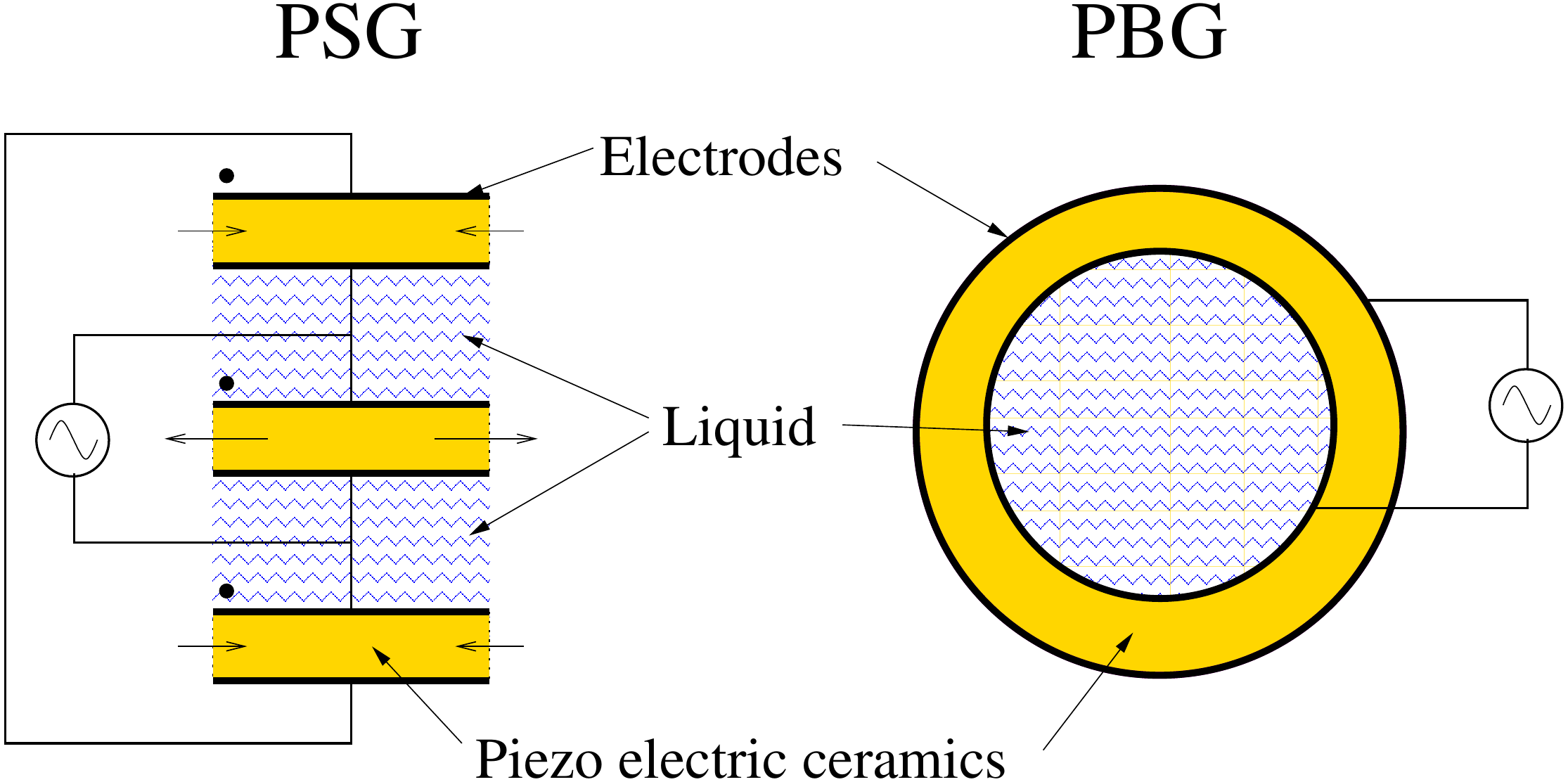}
  \end{center}
  \caption{\label{PGsetups}Schematic drawing of the PSG and PBG. In
    the PSG three electrode-covered piezo-electric ceramic discs are
    mounted in a layer construction; the liquid is loaded into the
    gaps between the discs. In the PBG, a spherical shell of
    piezo-electric ceramic material is filled with the liquid via a
    hole drilled in the ceramics. A liquid reservoir (not shown here)
    is attached above the hole, allowing extra liquid to be drawn in
    as the temperature is lowered.}
\end{figure}

Both the piezo-electric shear modulus gauge (PSG) and the
piezo-electric bulk modulus gauge (PBG) methods are based on the
piezo-electric effect, i.e., the conversion of electrical to
mechanical energy. These methods have been documented in detail by
Christensen and Olsen in Refs. \onlinecite{sChristensen1995} and
\onlinecite{sChristensen1994b}, respectively. In the following we
briefly sketch the steps in modeling of the devices, which allow us to
deduce a mechanical modulus from the electrical data.

\subsubsection{Piezo-electric Shear Modulus Gauge}

The PSG is constructed of three electrode-covered piezo-electric
ceramic discs mounted in a layered construction, which prevents
unwanted bending of the discs and further has the advantage that it
can be mapped mathematically to a one-disc system involving a fixed
wall \cite{sChristensen1995}.

The liquid is loaded into the $0.5$~mm gap between the discs
(Fig.\ \ref{PGsetups}). Depending on the polarity of the discs as
compared with the direction of an applied electric field, the discs
expand or contract in the radial direction. Electrically, the middle
disc is connected in parallel with the two outer discs in series as
shown in Fig.\ \ref{PGsetups}. Here, the small dots indicate the
polarity of the piezo-electric discs; thus when an electric field
is applied, the middle disc moves in opposition to the two outer
discs.  With this construction the gap between the discs is
field free, and the liquid is subjected to a purely mechanical
perturbation.

The capacitance of each disc depends on its strain state, so if the
liquid is partially clamping the disc (thus hindering its motion), the
measured capacitance is lower than that of freely moving discs. By a
precise measurement of the electrical capacitance of the PSG one can
obtain the stiffness of the liquid in contact with the disc. In other
words, knowing the exact relationship between the two, we can convert
the electric impedance into the shear modulus.

The elasto-electric compliance matrix describes the connection between
the components of the stress $\sigma_{ij}$ and strain $\epsilon_{ij}$
tensors and the electrical field of the piezo-electric material. The
equations describing a axially polarized ceramic can be split into
four independent parts, the relevant components of which can be
reduced to the following
\begin{equation}\label{eomatrix}
  \left( \begin{matrix}
      \sigma_{rr} \\  \sigma_{\phi\phi} \\ D_z\end{matrix}
  \right) = \left( \begin{matrix}  c_{11} & c_{12} &
      -e_{13} \\ c_{12} & c_{11} &  -e_{13} \\ e_{13} & e_{13} &
      \varepsilon_{33}^S \end{matrix} \right) \left( \begin{matrix}
      \epsilon_{rr} \\  \epsilon_{\phi\phi} \\ E_z\end{matrix}
  \right)
\end{equation}
where $c_{11}$ and $c_{12}$ are elastic constants of the ceramic,
$\epsilon_{33}^S$ is the dielectric constant, and $e_{13}$ is the
coupling constant.

The measured capacitance $C_m$ of the disc can be found by integrating
the charge density $D_z$ and dividing by the voltage
\begin{equation}
  C_m = \frac{Q}{U} = \frac{\int_0^{r_0}  2\pi r D_z(r)\,dr}{\xi
  E_z},
\end{equation}
where the charge density $D_z$ is given by Eq.\ (\ref{eomatrix}) and
$\xi$ is the thickness of the disc. $D_z$ depends both on the strain
state and the applied electrical field $E_z$. Evaluating this integral
it is found that the capacitance is a function of the radial
displacement at the edge of the disc $u_r(r_0)$
\begin{equation}
  C_m = Au_r(r_0)+B
\end{equation}
where $A$ and $B$ are known constants. It remains to determine the
displacement at the edge of the disc $u_r(r_0)$ as a function of
rigidity of the liquid. The displacement $u_r$ is found by solving the
radial equation of motion, which reduces to
\begin{equation}
  c_{11}\left(r^2(u_r'') +  u_r'-u_r  \right)  - \sigma_l
  \frac{r^2}{\xi} = -\omega^2r^2\rho u_r
\end{equation}
where the prime indicates the derivative with respect to $r$, $\xi$ is
the thickness of the disc and $\sigma_l$ is the tangential stress that
the liquid exerts on the disc. $\sigma_l$ is by definition
proportional the shear modulus of the liquid $\sigma_l= G(\omega)
u_r/d$, where $d$ is the thickness of the liquid layer (or
equivalently the distance between the discs), which is the quantity
relevant to determining the relaxation.

\begin{figure}
  \includegraphics[width=8cm]{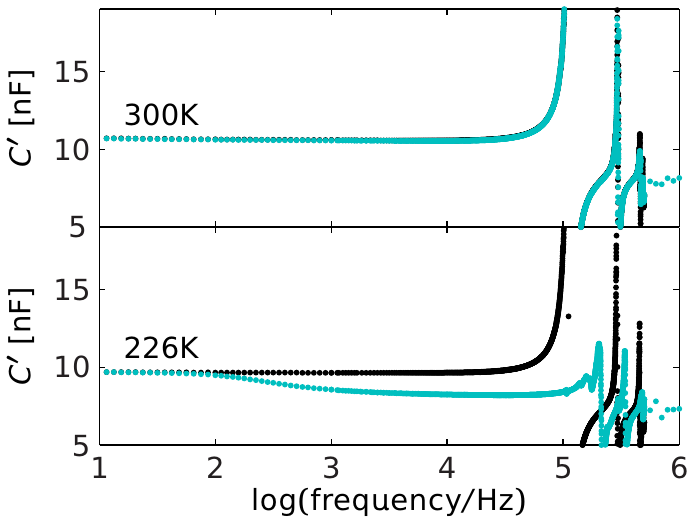}
  \caption{\label{rawshear}Raw data of the empty (black) and
    liquid-filled (blue) PSG at two different temperatures. At 300 K
    the presence of the liquid does not affect the signal because the
    liquid is quite fluid. At 226 K the liquid partially clamps the
    discs, which is manifested as a drop in capacitance in the
    quasi-static region and a shift of the resonances in the
    high-frequency region.}
\end{figure}

\begin{figure}
  \includegraphics[width=8cm]{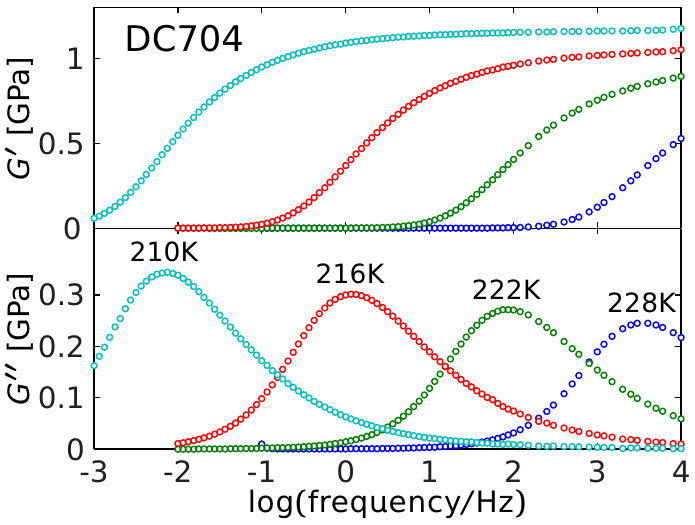}
  \caption{\label{sheardata}Real and imaginary parts of the shear
    modulus of DC704 at four temperatures approaching $T_g$.}
\end{figure}

Figure \ref{rawshear} shows the measured capacitance of the empty
(black trace) and liquid-filled (blue trace) PSG. At high temperatures
there is no influence from the liquid at these frequencies and the two
spectra are identical. The resonances in the spectrum are mechanical
resonances of the discs. At lower temperatures, the shear modulus of
the liquid increases and partially clamps the discs. This is observed
as a drop in the capacitance below the first resonances. We will refer
to the range of frequencies below the first resonance ($<100$~kHz) of
the system as the \emph{quasi-static} region. The liquid also
influences the positions of the overtones as compared with the spectrum
of the empty device. In the quasi-static region, the shear modulus is
found via the described inversion procedure. The inverted data, i.e.,
the inferred complex shear modulus, are shown in
Fig.\ \ref{sheardata}.

\subsubsection{Piezo-electric Bulk Modulus Gauge}

The PBG, which was also depicted in Fig.\ \ref{PGsetups}, consists of a
spherical shell of a piezo-electric ceramic material polarized in the
radial direction. The shell is covered by an electrode material both
on the inside and the outside. Applying an electric field to the
capacitor, which these electrodes constitute, deforms the ceramic
(expanding or contracting depending on the direction of the field) and
effectively changes the inner volume of the sphere.

A liquid inside the shell will oppose this deformation and thus change
the measured capacitance. The difference in capacitance between the
empty, freely moving shell and the partially clamped shell can be
related to the bulk modulus of the liquid. The deformation is radial.
An analysis of forced vibrations in a visco-elastic sphere shows
that in the low-frequency (quasi-static) region of the measurement
this corresponds to an isotropic compression of the liquid, while at high
frequencies it is a mixture of bulk and shear deformations.

In order to be able to fill the PBG with liquid, a hole is drilled in
the shell. A tube is attached over the hole. Filling this tube, as
well as the entire shell, allows the PBG to draw in extra liquid when
the liquid in the shell contracts during cooling. Thus the liquid
volume is constant throughout the duration of the measurement, i.e.,
at all temperatures.

\begin{figure}
  \includegraphics[width=8cm]{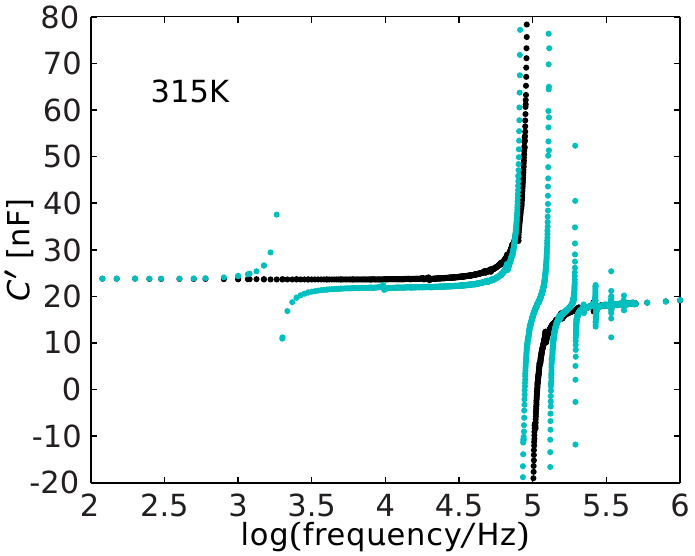}
  \caption{\label{raw_ex_dc704}Raw data of the empty (black) and the
    liquid-filled (blue) PBG. In the spectrum of the empty cell there
    is only one resonance (at $\sim 100$~kHz). There are a number of
    extra features that all come from the liquid. At low frequencies
    the capacitance of the two is the same. Around 1~kHz there is a
    small resonance, which comes from the liquid flowing in and out of
    the shell via a tube. Above that resonance the shell is partially
    clamped and the measured capacitance is reduced compared to that
    of the empty PBG. At frequencies above the first big resonance,
    the extra resonances are all due to standing waves in the liquid.}
\end{figure}

\begin{figure}
  \includegraphics[width=8cm]{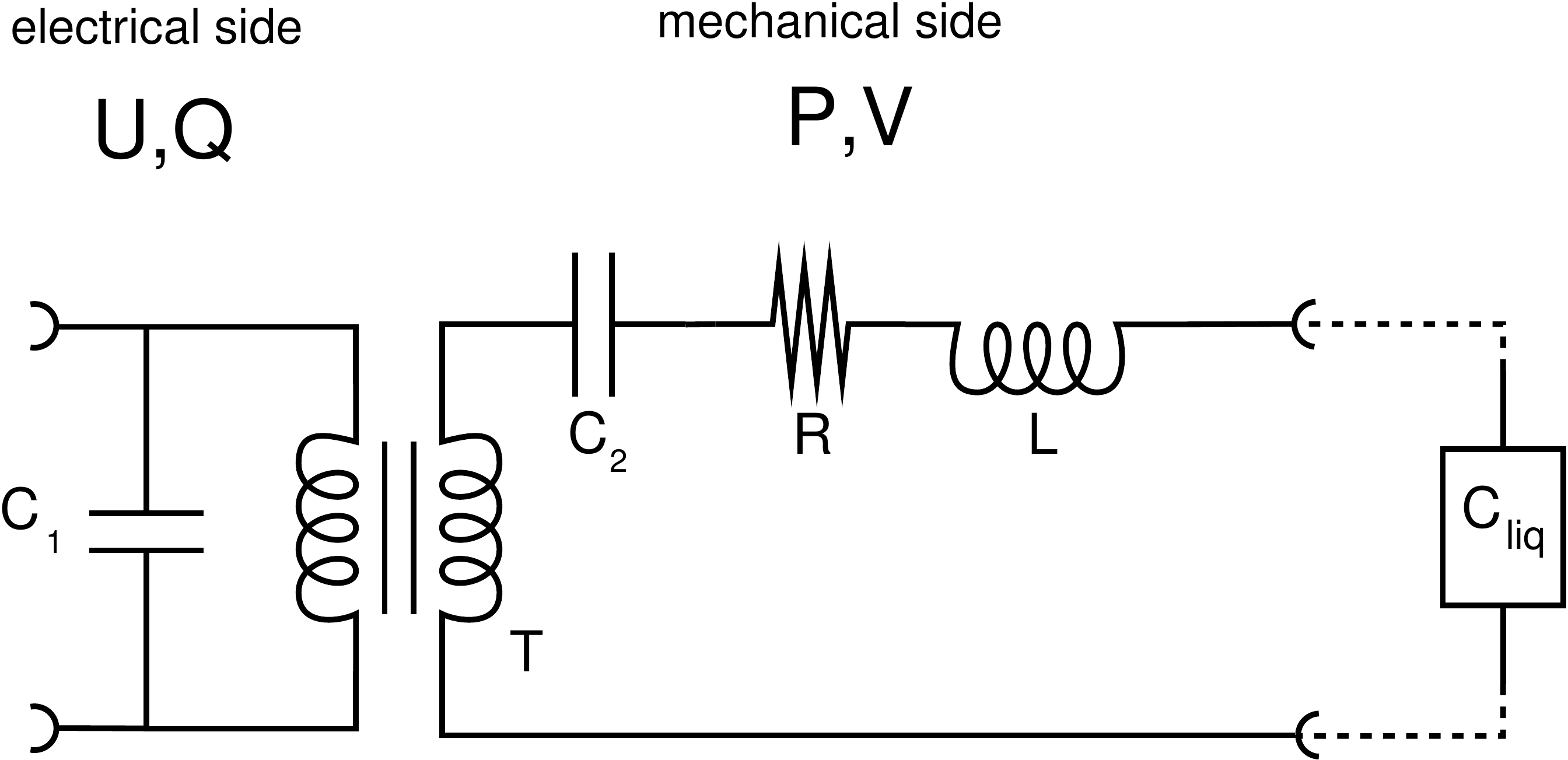}
  \caption{\label{PBG_model}Electrical equivalent model of the
    PBG. The transformer represents the conversion from electrical to
    mechanical energy in the piezo-electric ceramic. On the electric
    side of the network the capacitor is an actual capacitor that the
    electrodes constitute. On the mechanical side, an $LCR$ series
    models the inertial, elastic, and dissipative properties of the
    ceramic. When the PBG is empty the mechanical port is short
    circuited, allowing the shell to move freely. The liquid-filled
    PBG is modelled with an extra box $C_{\textrm{liq}}$ in series.}
\end{figure}

The modeling of the PBG is somewhat simpler than that of the PSG since
one can assume that the thickness of the ceramic is negligible in the
direction of its motion. Thus we can express the model in terms of an
electrical equivalent diagram, shown in Fig.\ \ref{PBG_model}, where
the conversion from electrical to mechanical energy is modeled by a
transducer $T_r$. On the electrical side of the diagram there is a
capacitor $C_1$ which corresponds to the actual capacitor constituted
by the two electrodes. On the mechanical side, the capacitor $C_2$
models the elastic properties of the ceramic, the inductor $L$ models
the inertance, and the resistor $R$ models the friction. The ``black
box'' $C_{\textrm{liq}}$ is the liquid capacitance which is what we want to
determine.

The capacitance of the model depicted in Fig.\ \ref{PBG_model} is given
by
\begin{equation}\label{pbgmath}
  C_m(\omega) = C_1 +  T_r^2 \frac{1}{\frac{1}{C_2} + i\omega R - \omega^2L +
    \frac{1}{C_{\text{liq}}}}.
\end{equation}
All the constants $C_1,C_2,L,R,T_r$ can be determined from a
measurement of the empty PBG. A subsequent measurement of the
liquid-filled PBG allows for the determination of the liquid's
mechanical stiffness $S_{\textrm{liq}}=1/C_{\textrm{liq}}$ by
isolating that term.

\begin{figure}
  \includegraphics[width=8cm]{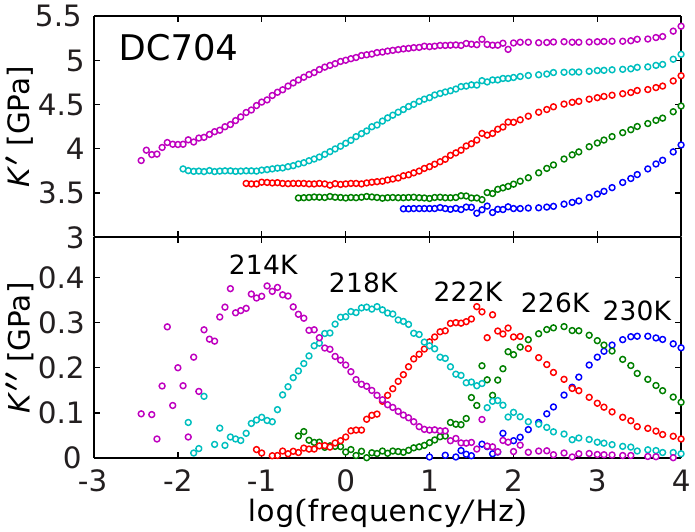}
  \caption{\label{bulkdata}Real and imaginary parts of the bulk
    modulus of DC704 for selected temperatures close to $T_g$.}
\end{figure}

Next, we wish to express the stiffness in terms of the elastic
moduli. The mechanical stiffness is $S_{liq}= \delta p / \delta V
\approx -\sigma_{rr}(r_0)/4\pi r_0^2u_r$. Solving the equation of
motion for forced vibrations in a visco-elastic sphere one finds the
displacement field $u_r$, which leads to the following expression for
the stiffness of the liquid \cite{sChristensen1994b}
\begin{equation}\label{stiffness}
\begin{split}
  S_{\textrm{liq}} & = \frac{1}{V}\left[K - M \left( 1 + \frac{1}{3} \frac{x^2
        \sin x}{x\cos x-\sin x}\right) \right].
\end{split}
\end{equation}
Here $V$ is the volume of the sphere, $K$ and $M$ are the (adiabatic)
bulk and longitudinal moduli, and $x=\sqrt{\rho/M}r_0\omega$. We have
$S_{\textrm{liq}} \rightarrow K/V$ as $\omega\rightarrow 0$, i.e., in the
quasistatic region we are measuring the bulk modulus. The inverted
complex bulk modulus data for DC704 are shown in Fig.\ \ref{bulkdata}.

\begin{figure}
  \includegraphics[width=8cm]{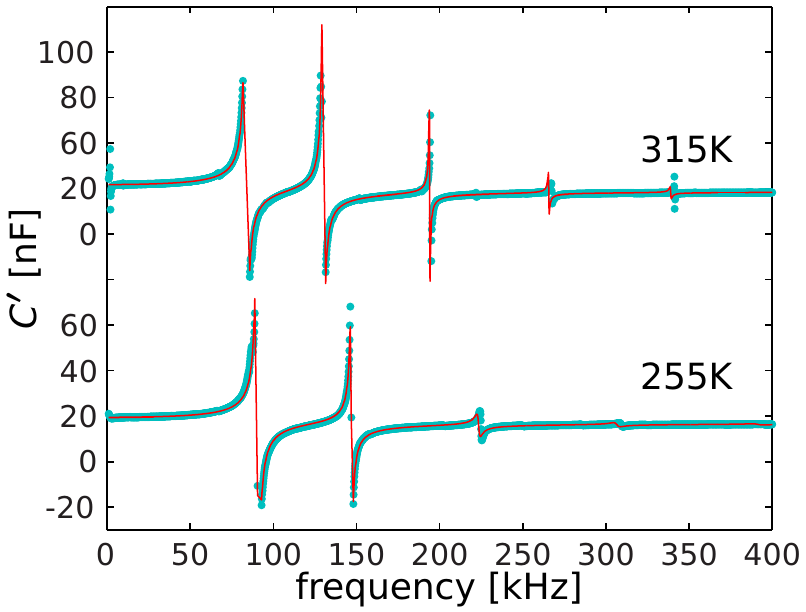}
  \caption{\label{bulkres_fitEx_dc704}Real part of capacitance in the
    resonance region of the PBG measurement. Data are blue dots and
    dashed lines are fits of a Eq.\ (\ref{stiffness}) in combination
    with the model in Eq.\ (\ref{pbgmath}).}
\end{figure}

At high frequencies we can fit Eq.~(\ref{stiffness}) in conjunction with
Eq.\ (\ref{pbgmath}) to the resonances, at least at high temperatures
where we can assume $K = K(\omega\rightarrow 0) = M(\omega\rightarrow
0)$ and the viscosity to be frequency-independent. Fits to data in the
resonance region are shown in Fig.\ \ref{bulkres_fitEx_dc704}.

\subsection{Nanosecond Acoustic Interferometry}\label{sec:NAI}

\begin{figure}
  \includegraphics[scale=0.32]{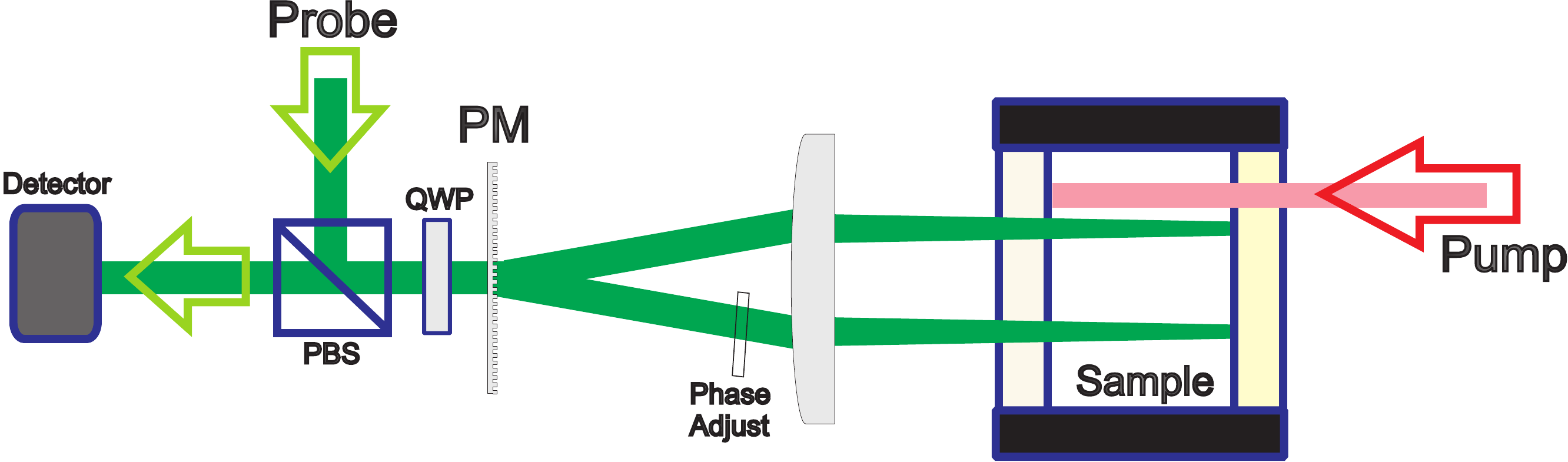}
  \caption{\label{fig:intersetup}Nanosecond acoustic interferometry
    setup.  A pump pulse is absorbed weakly in the sample, and sudden
    heating and thermal expansion generate a cylindrical acoustic wave
    radiating outwards from the pump beam. This acoustic pulse passes
    successively through each arm of a grating interferometer, and
    time-dependent changes in refractive index translate to
    time-dependent intensity of the interferometrically recombined
    signal.}
\end{figure}

In Nanosecond Acoustic Interferometry (NAI) acoustic wavepackets with
a distribution of frequency components in the low MHz range are
generated by a picosecond duration pump beam loosely focused in the
sample. Weak absorption leads to heating and thermal expansion of the
excited region, impulsively generating an outward propagating
cylindrical wave \cite{sPatel1981,sNeubrand1992}. This wave is
sequentially detected at the two arms of grating interferometer
\cite{sHess1996,sGlorieux2004} due to the phase difference created by
density-mediated changes in the refractive index.  The interferometer
arms are recombined at a diffraction grating, and we measure
time-dependent changes in the intensity of the resulting single beam
directed to a detector.

The setup is depicted in Fig.\ \ref{fig:intersetup}. To achieve
maximum sensitivity, the incoming continuous wave probe beam passes
through a polarizing beam splitting cube (PBS) and then a quarter
waveplate (QWP) to induce circular polarization. After passing through
the diffractive optic (binary phase mask pattern: PM), the $\pm 1$
diffraction orders are brought parallel and focused by a lens through
the sample cell to a dichroic mirror at the back of the sample, which
reflects the probe beams and transmits the pump.  The returning beams
recombine at the phase mask, pass again through the QWP, and the
resulting single beam is of correct polarization for transmission
through the PBS to a fast detector. A glass plate in the path of one
interferometer arm was rotated for relative phase control. Signals
were recorded at the relative phases $-\pi/2$ and $\pi/2$, midway
between maximum constructive and destructive interference, where the
induced phase change, and therefore the detected response, was
greatest \cite{sGlorieux2004}.

\begin{figure}
  \includegraphics[width=8.4cm]{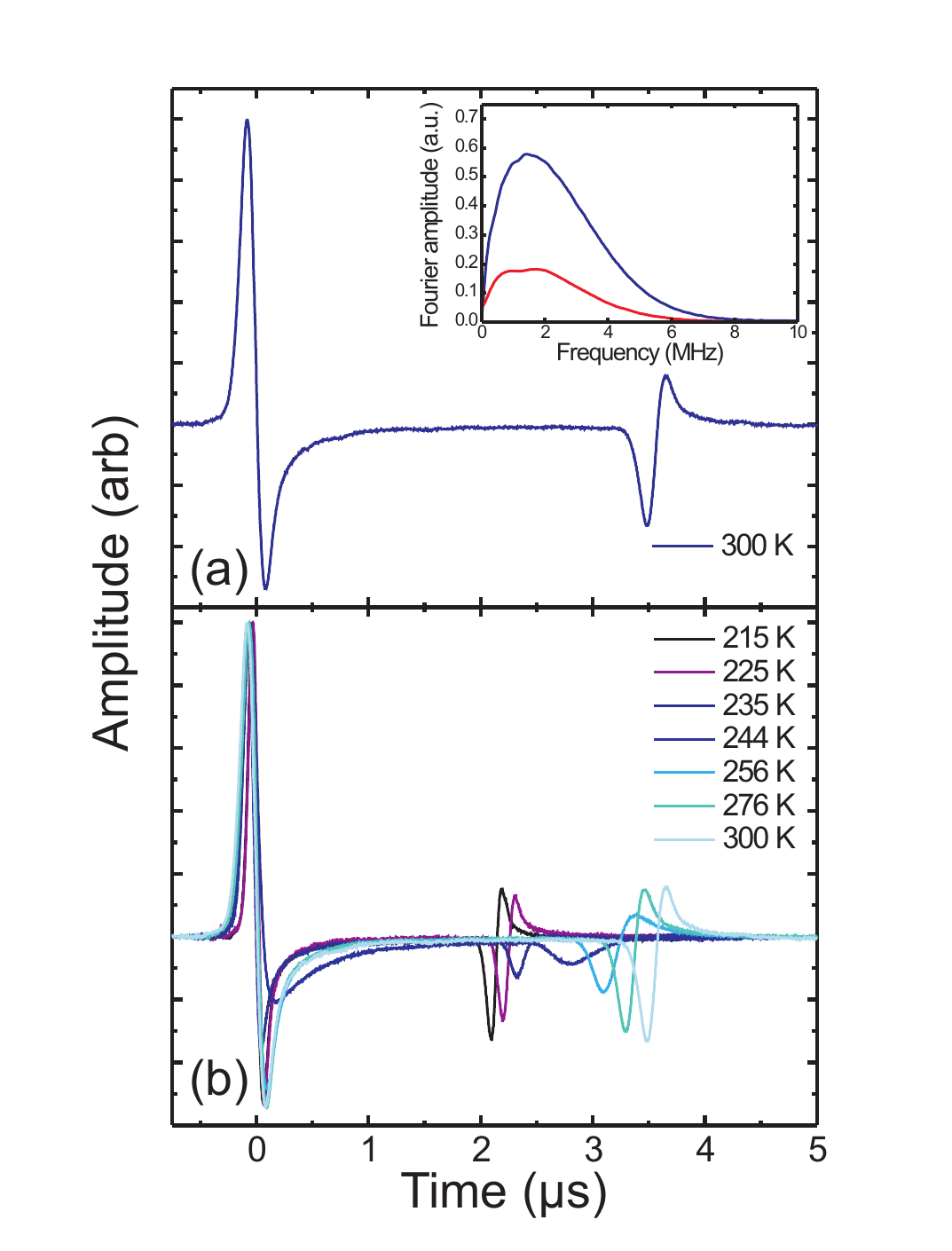}
  \caption{\label{fig:tdeptraces}(a) NAI trace recorded in DC704 at
    room temperature. A pump pulse is absorbed in the sample,
    generating a cylindrical wave radiating outwards. We observe the
    acoustic pulse passing first through one probe point and then
    through the second, where the amplitude reduction is clearly
    observed.  The inset shows the Fourier amplitude of the pulse at
    the first (blue curve) and second (red curve) points. (b)
    Nanosecond acoustic interferometry measurements for different
    temperatures.}
\end{figure}

Figure \ref{fig:tdeptraces}(a) shows a recorded trace of an acoustic
pulse generated in liquid DC704 passing through the first probe arm at
$t=0$ and the second at later times.  The inset shows the Fourier
amplitude of the pulse at each point, where the attenuation of the
wave is apparent through the relative change in spectral amplitude. In
Fig.\ \ref{fig:tdeptraces}(b), we show traces recorded at a number of
different sample temperatures for a set of traces normalized to the
maximum amplitude.

The effect of the coupling of structural relaxation to the acoustic
waves is twofold. First, we observe that as temperature is decreased,
the wavepacket arrives at the second probe point more quickly,
corresponding to an increasing speed of sound at lower temperatures.
Second, we note that the amplitude of the wavepacket has a strong
temperature dependence; acoustic damping is relatively low at high
temperatures in the liquid state, but as temperature is reduced,
strong damping is evident due to the alpha-relaxation. As
temperature is decreased further, the alpha-relaxation shifts to
lower frequencies and the damping is reduced at the acoustic
frequencies of our measurement where we see a solid-like response. We
note that even in the absence of acoustic damping, the amplitude of
the pulse at the second probe point decreases solely due to the
divergent nature of the cylindrical wavepacket, which is accounted for
in the analysis below.

The velocity and attenuation of the excited frequency components are
determined using the Fourier transform of the acoustic pulse at each
detection point according to
\begin{equation}
  \tilde{f}(\omega) = \mathcal{F}(f(t)) = \frac{1}{\sqrt{2\pi}}
  \int_{-\infty}^{\infty}f(t)e^{-i\omega t}dt\,.
\end{equation}
Using the shift theorem, i.e., that $\mathcal{F}(f(t-a))=e^{-i\omega
  a}\tilde{f}(\omega)$, and the distance $d$ between interferometer
probe arms, we can recover the speed of sound for the excited
frequency components via the phase. The acquired phase $(\phi)$ of the
Fourier transform is $\omega a$, and thus by taking the derivative of
the phase $(\partial_\omega\phi)$ with respect to frequency, we
recover the frequency-dependent shift in time $a(\phi)$. Using the
Fourier transforms of the wavepacket at the first probe point and
second probe point, we can determine the frequency-dependent sound
speed via
\begin{equation}\label{eq:velo}
  v(\omega)=\frac{d}{\delta t(\omega)}=\frac{d}{\partial_\omega
    \phi_1 (\omega)-\partial_\omega \phi_2 (\omega)}.
\end{equation}
where the subscripts 1 and 2 refer to the first and second probe point,
respectively. In practice, the amplitude of the phase recovered from a
numerical Fourier transform may depend on the number of time points in
the time-domain trace, and so we needed to multiply Eq.\ \ref{eq:velo}
by a constant calibration factor.  All recorded traces contained an
identical number of time points, and therefore a single calibration
factor for all analysis was sufficient.  The calibration factor was
picked to recover the sound speed of the solid phase determined at low
temperatures from the PBG/PSG and ISS data; this calibration factor
also recovered the liquid sound speed at high temperatures (in
comparison to ISS data), and so we use it with confidence for all
temperatures.  All points presented here were also collected with the
same probe distance $d$, which can be determined using the grating
period and the lens focal distance according to
\begin{equation}
  d=2f\tan\left(\sin^{-1}\left(\frac{\lambda_p}{2\Lambda}\right)\right),
\end{equation}
where $f$ is the lens focal length, $\lambda_p$ is the probe
wavelength, and $\Lambda$ is the grating period. Using our
experimental parameters of a 532~nm probe wavelength, 85~mm lens focal
length, and $8.5~\mu$m grating period, this translated to $\sim 5$~mm
distance between probe beams which agreed with a physical measurement
of the distance.

The frequency-dependent acoustic attenuation $\gamma(\omega)$ can be
determined by the amplitude of the Fourier transform of the acoustic
pulse at each probe point according to the Beer-Lambert law as
\begin{equation}
  \alpha(\omega)=\frac{\ln
    (|\tilde{f}_2(\omega)|/|\tilde{f}_1(\omega)|)}{d}.
\end{equation}
Here $|\tilde{f}_i(\omega)|$ denotes the magnitude and $\phi_i$
denotes the phase of the complex Fourier transform of the acoustic
pulse at each point. As we used a round excitation spot to generate
cylindrical acoustic waves with a well-defined $r^{-1/2}$ reduction in
amplitude, we account for the correction due to the acoustic wave
divergence by the constant $A$, which is given in terms of $d$ and the
distance between pump spot and probe $d_p$ as
\begin{equation}
  A=\sqrt{\frac{d_p+d}{d_p}}.
\end{equation}
We note that the attenuation is given here in units of inverse
distance, whereas the damping rate is in units of inverse time;
conversion between the two can be performed using the speed of sound.

For the broadband spectral analysis we can then construct the complex
acoustic modulus by
\begin{equation}
  M'(\omega) = \rho v^2(\omega) \frac{1 - \left( \alpha(\omega)
      v(\omega) / \omega \right)^2}{\left[1 + \left(\alpha(\omega)
        v(\omega) / \omega\right)^2 \right]^2}
\end{equation}
\begin{equation}
  M''(\omega)=2\rho v^2(\omega) \frac{\alpha(\omega) v(\omega) /
    \omega}{ \left[ 1 + \left( \alpha(\omega) v(\omega) / \omega
      \right)^2\right]^2},
\end{equation}
which were the quantities used in our analysis below.

Determination of very weak attenuation coefficients becomes unreliable
for a particular frequency when the signal-to-noise ratio in the
Fourier domain is too low.  In practice this was observed when, for a
particular frequency in the Fourier domain, the ratio of the magnitude
of the difference in acoustic signal amplitudes between the first and
second detection points (the amplitude at the second point being
corrected by the factor A) and the magnitude of the noise at the
second detection point was lower than approximately three to one.  For
such low attenuation rates, the imaginary part of the modulus could
not be determined.

\subsection{Impulsive stimulated scattering}\label{sec:ISTS}

\begin{figure}
  \includegraphics[scale=0.32]{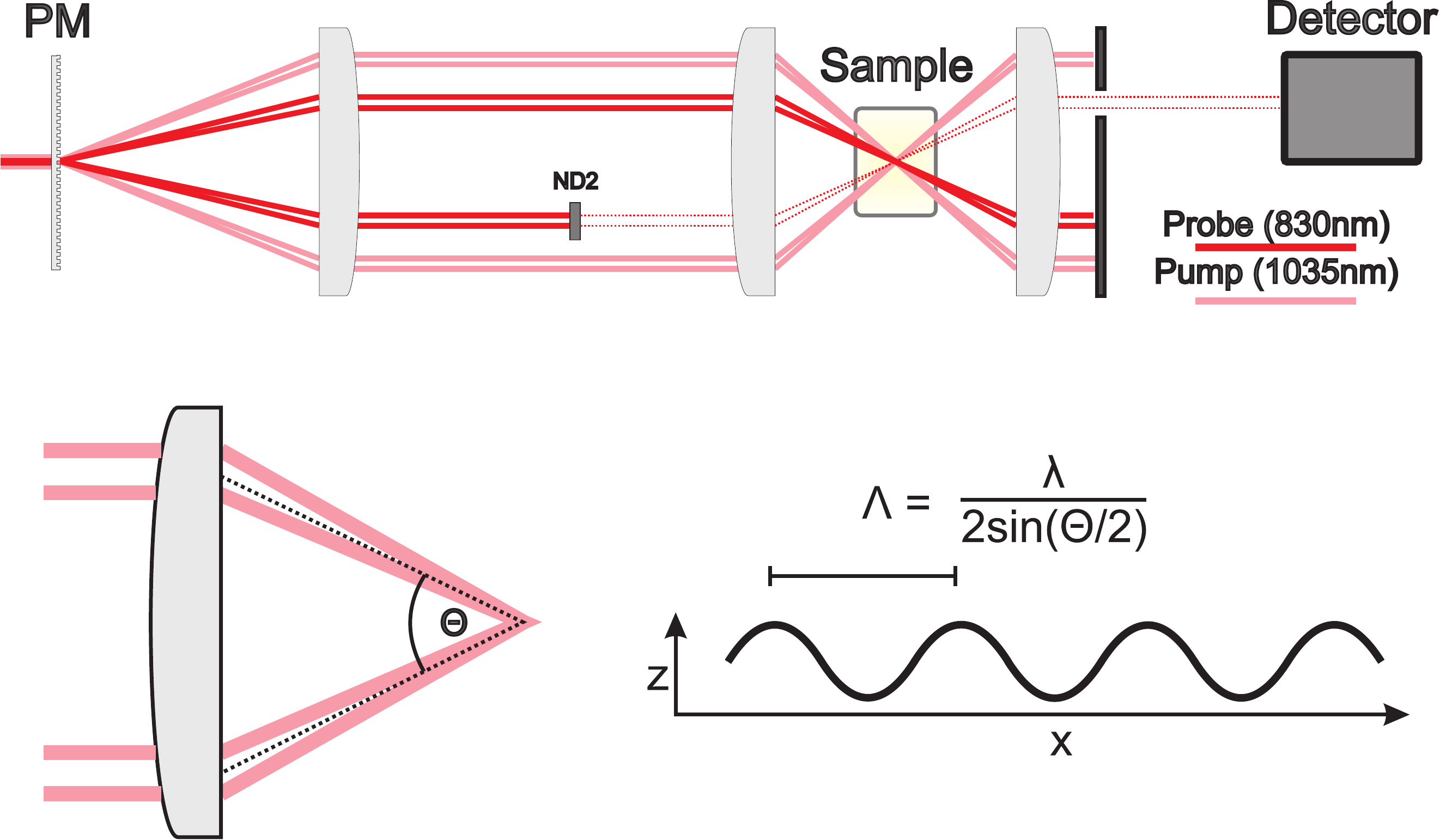}
  \caption[Schematic illustration of the ISS setup]{\label{fig:expt}
    Schematic illustration of the ISS setup. Both the pump and probe
    beams are incident on the phase mask (PM) and their $\pm$1
    diffraction orders are recombined at the sample at an angle
    $\theta$.}
\end{figure}

In an impulsive stimulated scattering (ISS) experiment
conducted in a heterodyned four-wave mixing geometry, light from a
pulsed laser is incident on a diffractive optical element, typically a
binary phase mask (PM) pattern, and split into two parts ($\pm 1$
diffraction orders; other orders are blocked), which are recombined at
an angle $\theta$ as depicted in Fig.\ \ref{fig:expt}. The crossed
excitation pulses excite an acoustic wave with wavelength $\Lambda$
given by the interference or "transient grating" period
\begin{equation}
  \Lambda=\frac{\lambda_e}{2\sin{\theta/2}}
\end{equation}
where $\lambda_e$ is the excitation laser wavelength. Probe light (in
the present case from a CW diode laser) is also incident on a phase
mask pattern (the same one or another with the same spatial period)
and split into two parts, which are recombined at the sample to serve
as probe and reference beams. The signal arises from diffraction of
probe light off the acoustic wave and any other spatially periodic
responses induced by the excitation pulses. The diffracted signal
field is superposed with the reference field for heterodyned
time-resolved detection of the signal, which typically shows damped
acoustic oscillations from which the acoustic frequency $\omega$ and
damping rate $\Gamma$ can be determined.

In an experiment where all four beams share common polarization V
(denoted VVVV), the action of the excitation pulses is twofold. First,
depending upon the absorbance at the pump wavelength, a fraction of
the light is absorbed into the sample and this energy is very rapidly
converted into heat. Sudden thermal expansion launches
counter-propagating acoustic waves with the acoustic period
$\Lambda$. In addition to acoustic oscillations, the signal may also
show slower, nonoscillatory density responses and thermal diffusion
from which complex structural relaxation dynamics and thermal
diffusivities may be determined \cite{sYang1995}. This excitation
mechanism is labeled Impulsive Stimulated Thermal Scattering (ISTS)
and is the dominant mechanism in the VVVV measurements.

The second excitation channel arises from the electrostrictive work
done on the liquid by the V-polarized excitations pulses. The electric
field of the interference maxima both induces a polarization and does
compressive work on the induced dipoles, resulting in impulsive
excitation of counterpropagating longitudinal acoustic waves of
wavelength $\Lambda$ even in the absence of optical absorption. This
excitation mechanism is termed impulsive stimulated Brillouin
scattering (ISBS). As the force scales with the gradient of the light
intensity, the efficiency of ISBS excitation scales as the acoustic
wavevector $q=2\pi/\Lambda$. Therefore this mechanism becomes more
important as the scattering wavevector is increased.

The excitation pulses may also induce molecular orientational
responses that can contribute to signal, analogous to depolarized
quasielastic scattering \cite{sSilence1992,sHinze2000}, as well as
contributions due to flow. In order to avoid this complication,
careful selection of the probe polarization relative to the grating
wavevector direction was used to eliminate these contributions to
signal.

The signal traces are fit to the model function~\cite{sYang1995}
\begin{multline}\label{eq:5gsol}
  A(e^{-t/\tau_{th}}-e^{-\Gamma_A t}\cos(\omega_A
  t))+B(e^{-t/\tau_{th}}-e^{-(t/\tau_s)^n})\\
  +Ce^{-\Gamma_A t}\sin(\omega t)
\end{multline}
where $A$ and $B$ are ISTS amplitudes, $C$ is the amplitude for the
ISBS signal, $\tau_{th}$ is the thermal decay time, and $\tau_s$ is
the characteristic structural relaxation time stretched by the
exponent n. Finally, $\omega_A$ is the observed acoustic
frequency, and $\Gamma_A$ is the acoustic damping rate. In regimes of
extremely low damping, the effect of finite pump and probe spot sizes
was explicitly taken into account using the results of Yan and
Nelson~\cite{sYan1987a} via a multiplication of the acoustic damping
term by a factor $e^{-\sigma(\omega t/q)^2}$, where $\sigma$ is given
terms of the spot sizes of the probe beam $\sigma_p$ and the
excitation beam $\sigma_e$ as
\begin{equation}
  \sigma=\frac{\sigma_{p}+2\sigma_e}{2\sigma_e\left(\sigma_p+\sigma_e\right)}.
\end{equation}

\begin{figure}
  \centering
  \includegraphics[scale=1]{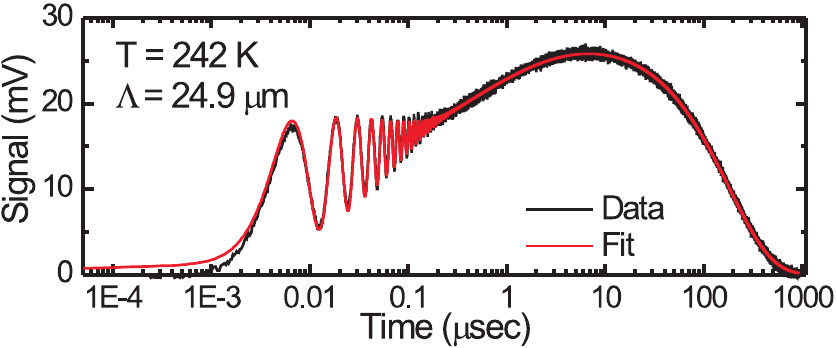}
  \caption[Longitudinal Signal with fit]{\label{fig:dc704lsig}ISTS
    signal in DC704 at 242K, 24.9~$\mu$m grating spacing. At short
    times, there is a damped acoustic oscillation due to
    counterpropigating acoustic waves. At later times, there is a slow
    rise due to the alpha-relaxation dynamics revealed though
    time-dependent thermal expansion. The signal eventually decays due
    to thermal diffusion of the thermal excitation grating.}
\end{figure}

Figure~\ref{fig:dc704lsig} details a representative trace with
$\Lambda=24.9~\mu$m which shows excellent agreement with the model
function, Eq.~(\ref{eq:5gsol}). These data, plotted on a logarithmic
time scale to cover signal extending over 6 decades, were taken in the
regime where the alpha-relaxation dynamics extend to time scales
significantly longer than that of the acoustic response, resulting in
insignificant acoustic damping and slow components of thermal
expansion that are observed directly in the data. This slow component
is represented by the stretched exponential function in the
time-domain signal function of Eq.~(\ref{eq:5gsol}). In principle this
slow signal reveals structural relaxation dynamics in the
0.1-100~$\mu$s temporal range and we could equate $\tau_s =
\tau_\alpha$. However, in addition to structural relaxation dynamics,
this component of the signal includes heat capacity relaxation
dynamics that may yield a complicated time-dependent evolution of the
sample temperature at the heated interference maxima. We expect that
this is a small contribution to the signal, but because we are not
able to determine it through independent measurement, we did not use
the slow component of our ISS data or the stretched exponential fits
to them in construction of our modulus curves.

\begin{figure}
  \centering
  \includegraphics[scale=0.95]{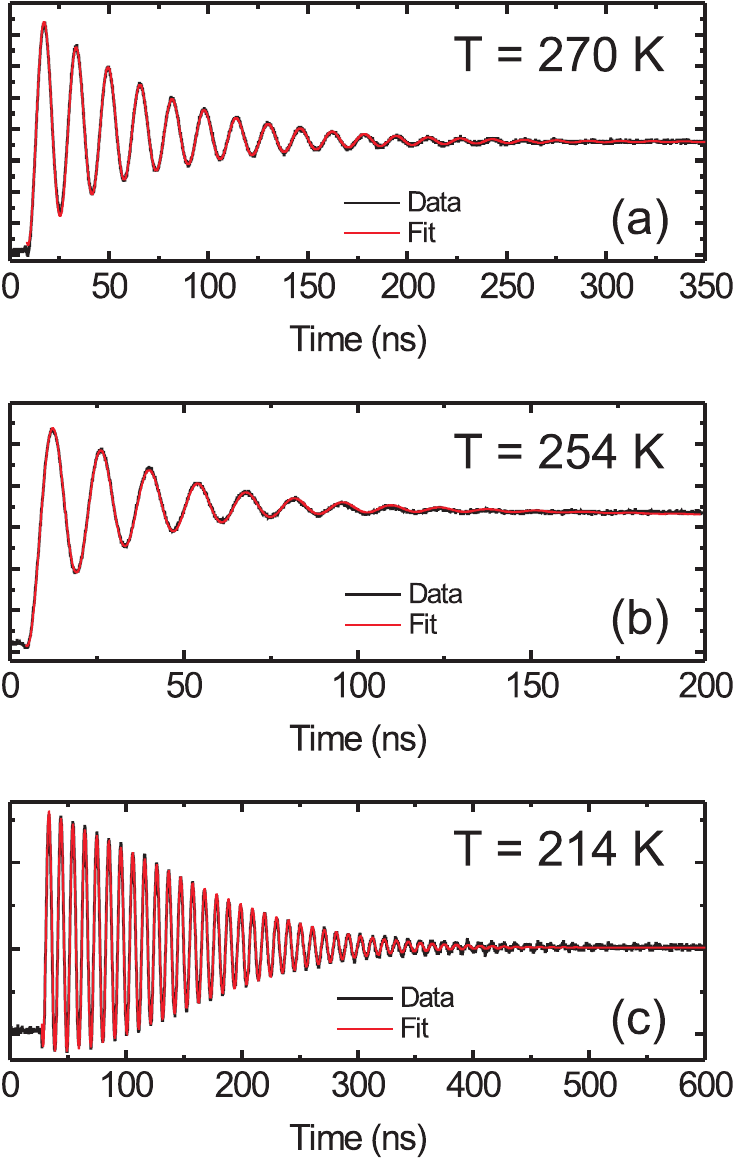}
  \caption{\label{fig:longfig2}ISTS signal in different temperature
    regimes using grating fringe spacing 24.9~$\mu$m. Fits are
    performed using Eq.~\ref{eq:5gsol}. In (a), the sample is in the
    liquid state and damped acoustic oscillations are observed on the
    sub-$\mu$s timescale. (b) As the sample is cooled, both the
    frequency and damping rate increase due to the presence of the
    alpha-relaxation dynamics at the acoustic frequency. (c) Near
    $T_g$, the acoustic frequency is greatly increased and the damping
    rate decreased due to the solid characteristic of the material.}
\end{figure}

For this study we focused exclusively on the acoustic
contribution. Figure~\ref{fig:longfig2} shows three more
representative traces of ISTS data recorded with 24.9~$\mu$m grating
spacing. In all traces there are oscillations due to
counterpropagating acoustic waves. The data in (a) are taken in the
high-temperature liquid state, showing weak damping of the acoustic
wave. In the regime where the alpha-relaxation is on the time scale
of an acoustic period, mechanical energy is quickly dissipated into
structural relaxation and the acoustic signal is strongly damped, as
shown in (b). Finally, when the liquid is cooled into the very viscous
state as in (c), the damping rate is observed to decrease
significantly. Here the effects of finite beam spot sizes lead to
non-exponential decay, which is modeled by the correction term
described above.

The transient grating spacing (\emph{i.e.}, wavevector magnitude $q$)
can be varied by changing the phase mask pattern to make the
excitation pulses cross at different angles in the sample, and new
data can be recorded which will reveal different acoustic frequencies
and damping rates as functions of temperature. Using these acoustic
parameters determined at many wavevectors at a common temperature, the
complex frequency-dependent longitudinal modulus
$\hat{M}(\omega)=M'(\omega)+iM''(\omega)$ at that temperature can be
determined from
\begin{eqnarray}
  M'(\omega)&=&\rho\frac{\omega^2-\Gamma^2}{q^2}\\
  M''(\omega)&=&\rho\frac{2\omega\Gamma}{q^2}.
\end{eqnarray}

The excitation beams were focused to a spot 2.5~mm in the grating
dimension and 100~$\mu$m in the perpendicular dimension so that the
acoustic waves would have many periods and the decay of signal would
be due primarily to acoustic damping rather than propagation away from
the excitation and probing region of the sample. In regions of
extremely low damping the effects of finite spot size were taken into
account, as detailed above. Damping rates as low as 2 $\mu
\textbf{s}^{-1}$ could be measured reliably.

The probe was focused to a spot of 1~mm in the grating dimension by
50~$\mu$m in the perpendicular dimension. We used a common phase mask
optimized for diffraction into $\pm$1 orders at 800~nm for both pump
and probe beams. The local oscillator was attenuated by a factor of
$10^{-3}$. Approximately $30\%$ of the pump power was lost into zero
order with this configuration, but the pump intensity still had to be
reduced significantly to avoid unwanted nonlinear effects. The signal
was collected with a fast amplified photodiode with 3~GHz bandwidth
and processed in a 4~GHz bandwidth digitizing oscilloscope. When
slower signals were studied, we used a New Focus Model 1801-FS
detector with a bandwidth of DC to 125~MHz. Depending upon
signal-to-noise ratios, signals from 2,000 to 4,000 repetitions of the
measurement were averaged for each data trace, with total data
acquisition times of less than a minute per trace.

The acoustic wavelength was calibrated through ISTS measurements in
ethylene glycol, for which the speed of sound is known accurately
\cite{ssilence_thesis}, and cross-checked with internal measurements of
DC704 at high temperature where the speed of sound is constant across
the range of measured acoustic wavelengths. This calibration was
double-checked after all of the data collection was finished, and the
variation in acoustic wavelength ranged from approximately 0.1\% to
1.8\%, with an average of 0.7\%. This determined our uncertainties in
the sound speeds, but did not affect the uncertainties in the damping
rate.

To build acoustic spectra, data were taken by fixing the sample
temperature and using every available phase mask pattern to provide
wavelengths in the range from 1.75~$\mu$m to 101~$\mu$m. These grating
spacings were 1.71~$\mu$m, 1.97~$\mu$m, 2.33~$\mu$m, 2.68~$\mu$m,
3.14~$\mu$m, 3.64~$\mu$m, 4.20~$\mu$m, 4.85~$\mu$m, 5.66~$\mu$m,
6.56~$\mu$m, 6.70~$\mu$m, 7.61~$\mu$m, 9.13~$\mu$m, 10.2~$\mu$m,
11.7~$\mu$m, 13.7~$\mu$m, 15.7~$\mu$m, 18.3~$\mu$m, 21.3~$\mu$m,
24.9~$\mu$m, 28.5~$\mu$m, 33.0~$\mu$m, 38.1~$\mu$m, 44.2~$\mu$m,
49.8~$\mu$m, 50.7~$\mu$m, 56.9~$\mu$m, 65.9~$\mu$m, 76.0~$\mu$m,
88.0~$\mu$m, and 101~$\mu$m. The procedure was repeated for different
temperatures until the desired range was covered.

\subsection{Time-Domain Brillouin Scattering ($\sim$3--23~GHz)}\label{sec:TDBS}

In time-domain Brillouin scattering (TDBS), acoustic wave
propagation at low GHz frequencies is monitored by an optical,
time delayed probe at a fixed wave vector and a complex frequency
is measured. The detection scheme is based on the coupling of
mechanical strain with laser light through the so-called
photoelastic or Brillouin effect.

\begin{figure}[th!]
  \includegraphics[width=.48\textwidth]{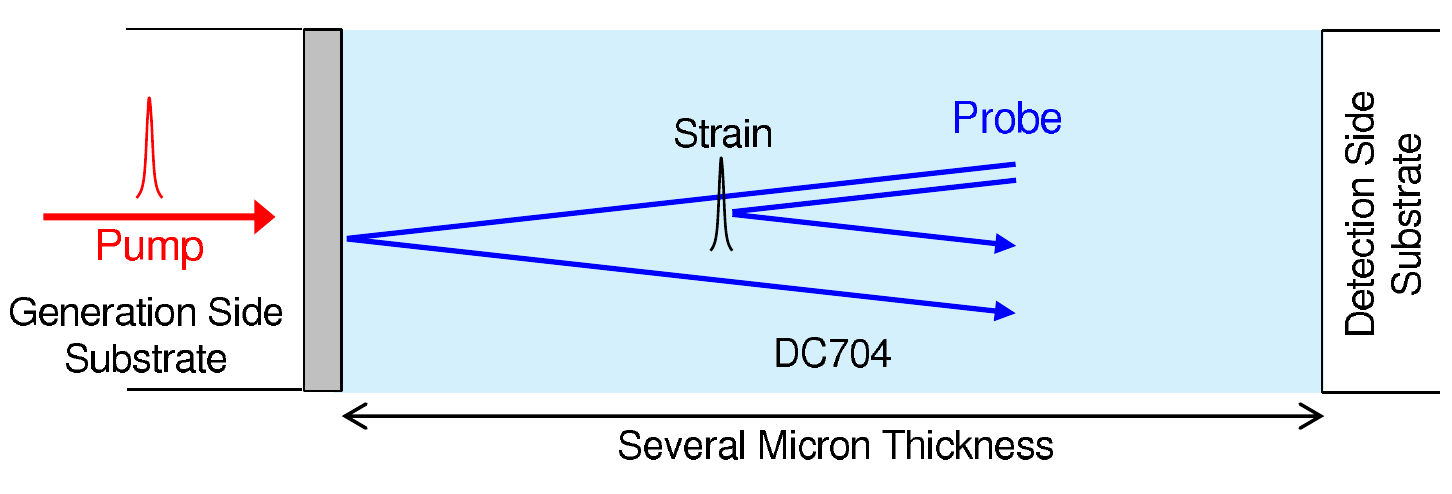}
  \caption{\label{fig:TDBS_sketch} Schematic diagram of time domain
    Brillouin scattering detection. An acoustic strain pulse is
    generated in a transducer film and propagates into an adjacent
    liquid layer.  Incident probe light reflects off of the transducer
    film, but also off the propagating strain pulse due to
    acoustically induced local changes in the refractive index, and
    interference gives rise to an oscillatory modulation of the
    overall reflected probe intensity.}
\end{figure}

A schematic illustration of this technique is shown in
Figure~\ref{fig:TDBS_sketch}. Acoustic waves are excited by a
sub-picosecond optical pump pulse which deposits energy in a
photo-acoustic aluminum transducer thin film. The subsequent sudden
thermal expansion launches an acoustic strain pulse $\eta(z,t)$ into
the adjacent liquid film. Probing of the propagation and attenuation
of the strain pulse away from the transducer is accomplished by a
second optical pulse mechanically delayed by a variable time $\Delta
t$ relative to the pump. The reflected probe light consists of a
number of superposing beams with a principal reflection from the
transducer and minor reflections from the strain pulse
\cite{sThomsen1986, sLin1991}. These reflections arise from the local
change in refractive index produced by the presence of the strain
pulse via the photoelastic or Brillouin effect. Depending on the
distance $d=c\Delta t$ of the strain pulse from the transducer film,
the reflected beams interfere either constructively or
destructively. This results in an oscillatory modulation of the total
reflected intensity as a function of time, which can be used to
determine the acoustic speed $c$ of the strain pulse in the sample
while the amplitude of the modulation can be related to the amplitude
of the strain.

\begin{figure}
  \includegraphics[width=.48\textwidth]{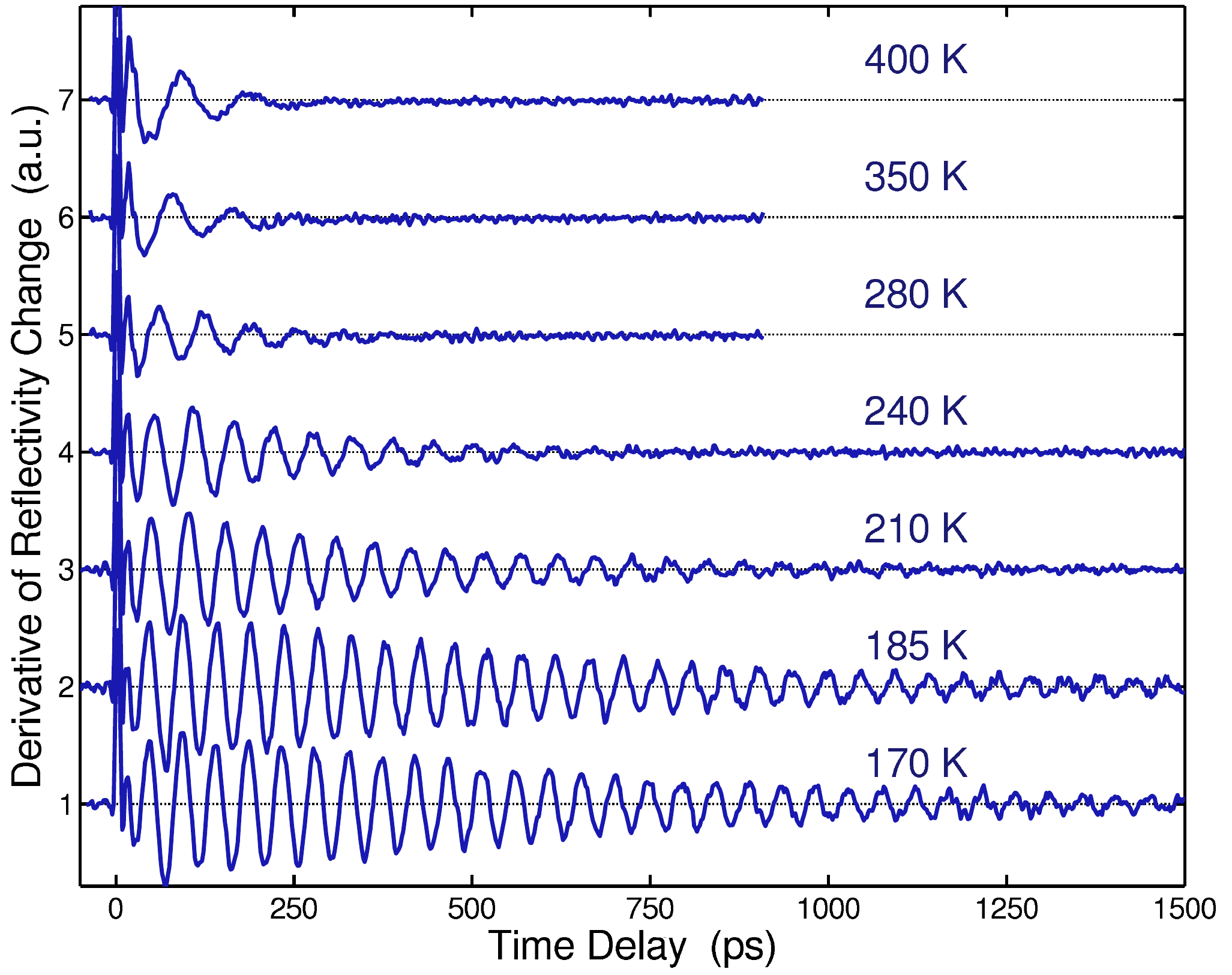}
  \caption{\label{fig:TDBS_rawdata}Selection of Brillouin scattering
    data in DC704 recorded at various temperatures. At the highest
    temperatures, the frequency is lower and the damping rate higher
    due to the presence of complex relaxation dynamics at the probed
    frequency. As the temperature is reduced, the frequency is
    observed to increase and the damping rate is observed to decrease
    significantly.}
\end{figure}

At normal incidence of the probe beam with respect to the transducer,
the measured Brillouin scattering frequency is
\begin{equation}
  f_{BS} = \frac{2nc}{\lambda} \; ,\label{eq:BS_frequency}
\end{equation}
where $n$ is the index of refraction and $\lambda$ is the optical
wavelength. A measurement of this frequency can thus give the sound
velocity when the refractive index of the sample is known. To
determine the acoustic attenuation rate at this frequency, the raw
data were fit to the time-domain form of a damped oscillator, i.e.,
$e^{-\Gamma t}\sin\left(\omega t\right)$.

For the purpose of compiling broad relaxation spectra, we recorded
time-domain Brillouin scattering data at many different temperatures
with probe light at the laser fundamental wavelength, centered at 790
nm, and frequency doubled light, centered at 395 nm. The sample was
constructed from a 20 nm aluminum transducer film on a sapphire
generation-side substrate and a many-micron thick liquid layer held by
another sapphire substrate. This construction allowed us to monitor
the propagation of the acoustic strain pulse until it was attenuated
to below our detection limit.

Selected raw data at several different temperatures are shown in
Figure~\ref{fig:TDBS_rawdata} and clearly illustrate the strong
acoustic attenuation at temperatures above the glass transition
temperature, $T_g=210$ K. The attenuation rate is observed to decrease
as the liquid is cooled into the glassy state. Results for the
extracted Brillouin scattering frequency and attenuation rate were
obtained by fitting a Lorentzian oscillator as described above.

\subsection{Picosecond Ultrasonic Interferometry
  ($\sim$25-120~GHz)}\label{sec:PUI}

\begin{figure}[ht!]
  \centering
  \includegraphics[width=.48\textwidth]{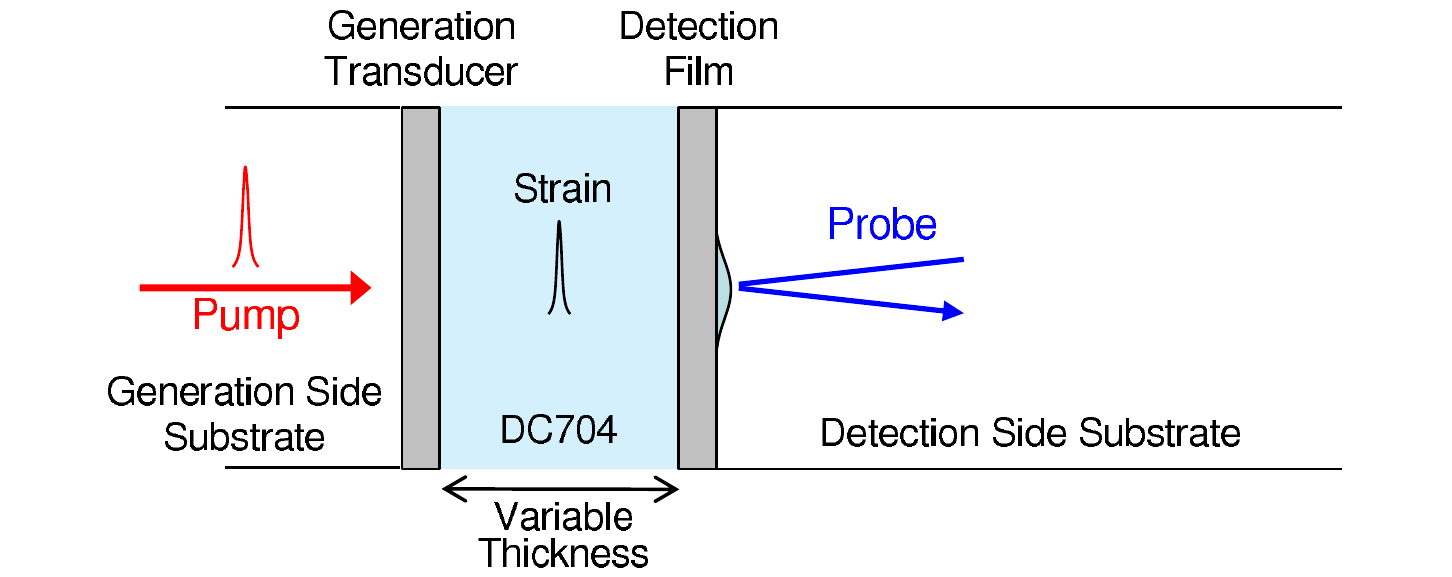}
  \caption{\label{fig:PUI_sketch}Sketch of picosecond ultrasonic
    interferometric experimental approach. Energy deposition by an
    ultra-short laser pulse in a transducer film resulted in sudden
    thermal expansion and launched a compressional strain pulse into
    and through the DC704 liquid sample. Its arrival in a detection
    film caused a small displacement of that film which was detected
    interferometrically.}
\end{figure}

Here we describe the manner in which a specialized picosecond laser
ultrasonic technique may be combined with interferometric detection
for probing longitudinal acoustic waves in the tens to hundreds of
gigahertz frequency range. This method, depicted in Fig.\
\ref{fig:PUI_sketch}, is performed in a similar sample geometry as in
time-domain Brillouin scattering. Single-cycle acoustic waves are
generated in a thin metal transducer film on one side of a liquid film
and their arrival and shape is measured by a Sagnac
interferometer~\cite{sHurley1999,sPerrin1999} at the receiver metal
film. In our analysis we compared transmitted strain pulses through
different liquid thicknesses with each other in order to extract the
liquid response at a fixed frequency.

\begin{figure}
  \centering
  \includegraphics[width=7.5cm]{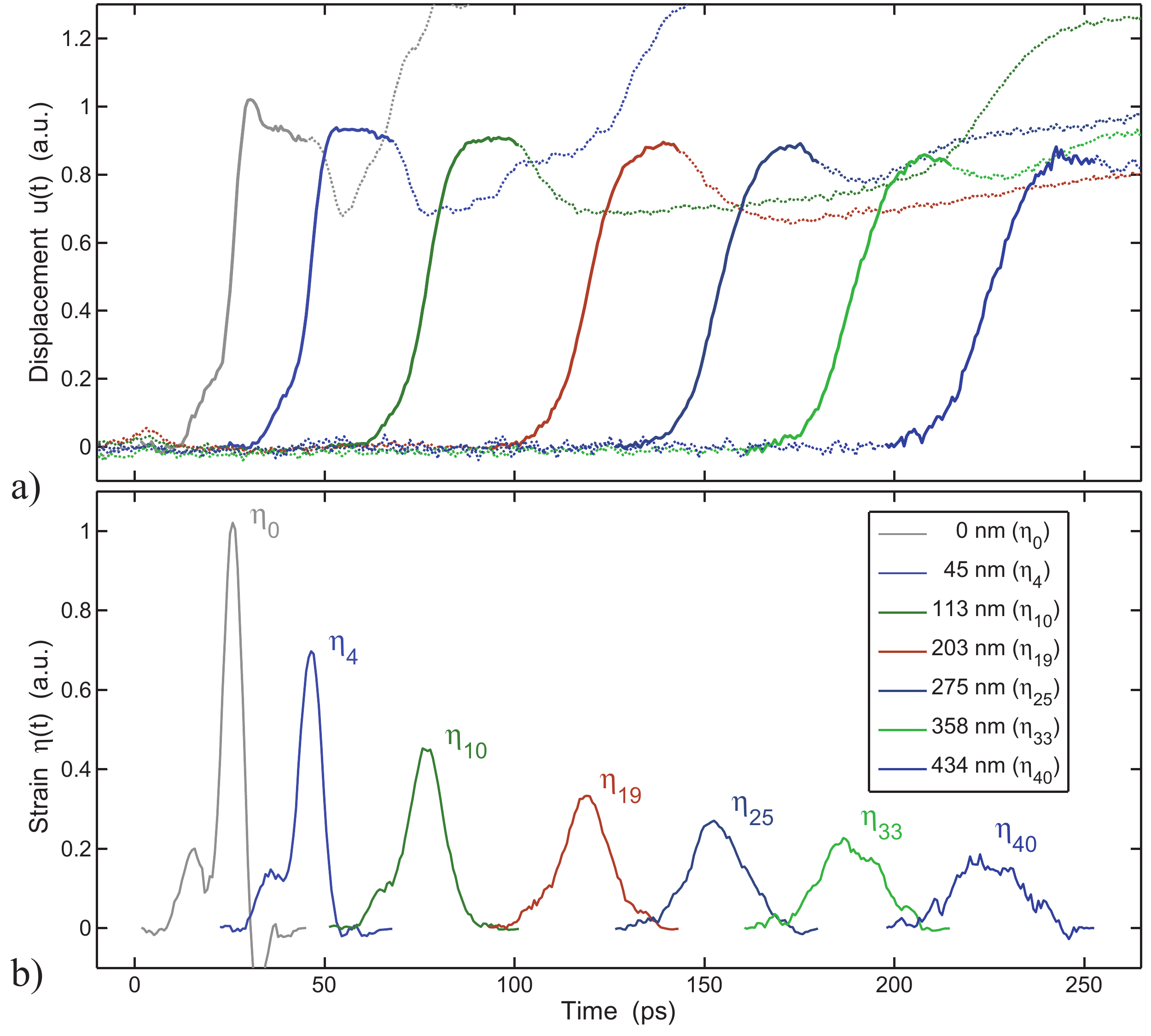}
  \caption{\label{fig:PUI_different_strains}Measured and normalized
    displacement data (a) and the derived strain (b) after
    transmission through different layer thicknesses of DC704 at
    200~K. The acoustic strain was produced by a 30~nm aluminum
    transducer film on a sapphire generation side substrate and
    detected interferometrically at the back side of a 100~nm aluminum
    film on a sapphire detection side substrate. Signals labeled
    $\eta_4$, $\eta_{10}$, $\eta_{19}$, $\eta_{25}$, $\eta_{33}$, and
    $\eta_{40}$ are from acoustic waves transmitted through
    progressively greater liquid thicknesses as indicated.}
\end{figure}

Our approach is to compare signals transmitted through different
thicknesses of sample material. In a manner similar to the NAI
experiment detailed in Sec.~\ref{sec:NAI}, we compare the amplitudes
and phases at a given frequency by Fourier transformation of the two
time-domain signals $h_1(t)$ and $h_2(t)$, which constitute two
signals transmitted through different sample thicknesses $d_1$ and
$d_2$~$>d_1$ (e.g., $\eta_4$ and $\eta_{10}$ in
Fig.\ \ref{fig:PUI_different_strains}).

Figure \ref{fig:PUI_different_strains}(a) shows a selection of the
measured displacements for several different thicknesses of DC704.
Typically, we measured the strain transmitted through about 50
different liquid thicknesses. A signal ($\eta_0$) from the part of the
sample assembly where the transducer films were in contact is added to
the plot. We did not use such signal amplitudes for our analysis since
the acoustic strain in this case did not experience exactly the same
conditions (due to the contact between the metal thin films) as in the
other cases, but the signal provided calibration for essentially zero
liquid thickness.

We assume that the displacement can be described by a plane wave
equation.  The time-derivative of the displacement $u(t,x)$ is then
proportional to the strain $\eta(t,x)$,
\begin{equation}
  \eta(t,x) = \frac{d}{dx} u(t,x) \;=\; \frac{1}{c} \frac{d}{dt}
  u(t,x) \;.
\end{equation}
where $c$ is the acoustic phase velocity (assuming $c$ is
non-dispersive).

Figure \ref{fig:PUI_different_strains}(b) shows the transmitted
strain through a selection of different liquid thicknesses, labeled
$\eta_4$, $\eta_{10}$, $\eta_{19}$, $\eta_{25}$, $\eta_{33}$, and
$\eta_{40}$. The shape of the strain after transmission through
different liquid thicknesses is simply related to a set of acoustic
parameters which includes the complex speed of sound or equivalently
the complex longitudinal modulus. After traveling through an
additional distance $\Delta d$ in the liquid, the Fourier domain
`output' strain $\tilde{\eta}_{out}(\omega)$ is related to the
`input' strain $\tilde{\eta}_{in}(\omega)$ by the complex
wavevector, $k(\omega)=\omega/c(\omega)$,
\begin{eqnarray}
  \tilde{\eta}_{out}(\omega) & = & e^{i\hat{k}(\omega) \Delta d} \;
  \tilde{\eta}_{in}(\omega) \;.
\end{eqnarray}
While the real portion of the wavevector shifts the phase of the input
strain, the imaginary component of the wavevector dampens the
amplitude. If we denote the transmitted strain as
\begin{eqnarray}
  \tilde{\eta}_{trans}(\omega) &=&
  \frac{\tilde{\eta}_{out}(\omega)}{\tilde{\eta}_{in}(\omega)} \;,
\end{eqnarray}
we can write the complex acoustic velocity as
\begin{eqnarray}
  c(\omega) &=& \frac{\omega}{k} \;=\;
  \frac{i\,\omega\,\Delta d}{\ln(\tilde{\eta}_{trans}(\omega))} \;.
\end{eqnarray}
From this, the complex acoustic modulus $M(\omega)$ and the
acoustic compliance spectrum $J(\omega)$ can be determined
directly from a broadband measurement of the transmitted strain:
\begin{eqnarray}
  M(\omega) &=& - \rho \, \left[ \frac{\omega\,\Delta
      d}{\ln(\tilde{\eta}_{trans}(\omega))} \right]^2 \;.
\end{eqnarray}

In order to account for different attenuation strengths at different
frequencies in the most reliable manner, we conducted our analysis on
data sets having about 50 strain profiles, each recorded at a
different liquid thickness using a fixed excitation pulse shape and at
a constant sample temperature.

\section{Analysis of Mechanical Spectra}\label{sec:phen-analys}

Below we provide a more in-depth discussion of some of the fitting
results presented in the main manuscript in the context of the mode
coupling theory.

\subsection{Time-temperature superposition}

The relaxation behavior at frequencies around the alpha-peak was
fitted by a stretched exponential equation
\begin{equation}\label{eq:KWW}
  \chi(t)=\chi_0+\Delta \chi \exp \left(-(t/\tau_{\alpha})^n\right)\,,
\end{equation}
where $\chi_0$ denotes the long-time limit of $\chi(t)$ and $\Delta
J\chi=\chi_0-\chi_\infty$ the relaxation strength.

The $M''(\omega)$ data were fitted to a numerical Fourier transform of
Eq.~(\ref{eq:KWW}) using a routine developed by
Wuttke~\cite{sWuttke2012}, yielding the fits to the data in Fig.\ 2(a)
of the main text. The compliance data $J'(\omega)$ and $J''(\omega)$
were also fit to the KWW Fourier transform using the same routine. In
order to demonstrate TTS, rescaling of the compliance data was
accomplished using our measured values for the limiting instantaneous
speed of sound ($c_\infty$)~\cite{sKlieber2013} whose temperature
dependence was found to be fit well by the form
\begin{equation}\label{cinf_DC704}
  c_\infty (T)=3840[\textrm{m/s}] - 6.9[\textrm{m/sK}]\times T \;
\end{equation}
to calculate $J_{\infty} = \sqrt{\rho/c_{\infty}}$ and the normalized
quantities in Fig.~3(a).

The low-temperature compliance data cover the alpha relaxation
spectrum quite thoroughly, and the alpha spectra were found to be fit
well by the KWW form with the stretching exponent $n = 0.5$. This
value was fixed for the fits at higher temperatures, leaving only the
three remaining fitting parameters $J_0$, $\Delta J$, and
$\tau_{\alpha}$. Restricting the number of fitting parameters made
possible fitting of spectra across the modest frequency gaps in our
spectra. We note that the measurements were typically conducted with
small temperature steps (usually $\sim 2$~K) but the measurements
conducted with different methods were not all made at exactly the same
temperatures. In order to show results spanning the entire frequency
range at selected temperatures (e.g., 215~K, 225~K, and 235~K in
Fig.~3a), we interpolated at the lower frequencies between data
measured at very nearby temperatures, e.g., 214~K and 216~K for the
results shown at 215~K. This procedure allowed us to reliably connect
results from our highest to lowest frequency ranges despite the number
of measurement methods involved.

\begin{figure}\centering
  \includegraphics[width=7.5cm]{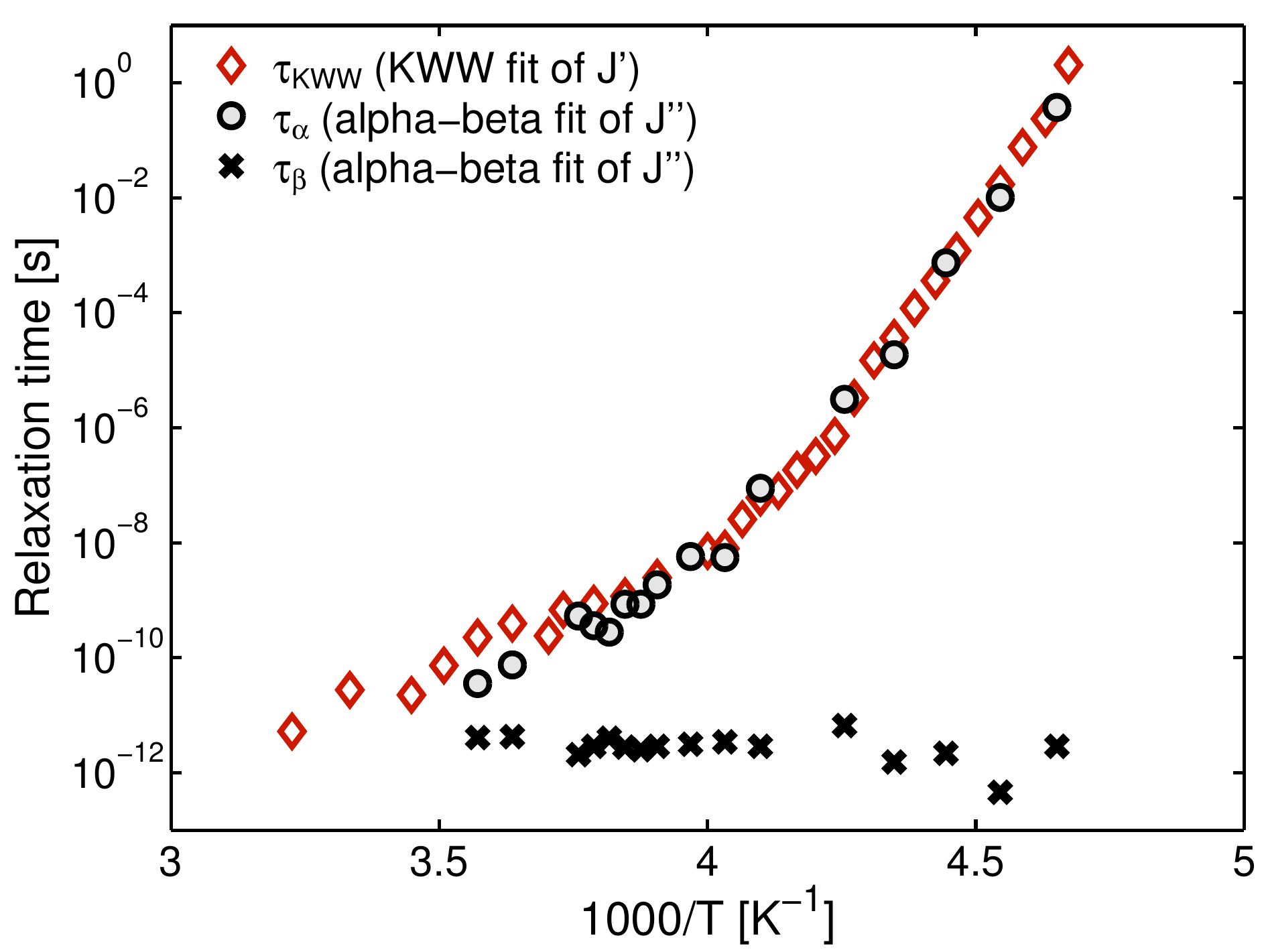}
  \caption{\label{fig:alphabeta} Relaxation times of the alpha and
    beta processes from fits of data to a phenomenological
    alpha-beta merging model expressed as a network diagram
    \cite{sSaglanmak2010,sJakobsen2012}. The alpha relaxation times of
    the alpha-beta model are markedly non-Arrhenius and agree well
    with the $\tau_{\alpha}(t)$ values of the stretched exponential fit. The
    beta relaxation time is roughly temperature-independent.}
\end{figure}

\begin{figure}
  \includegraphics[width=7.5cm]{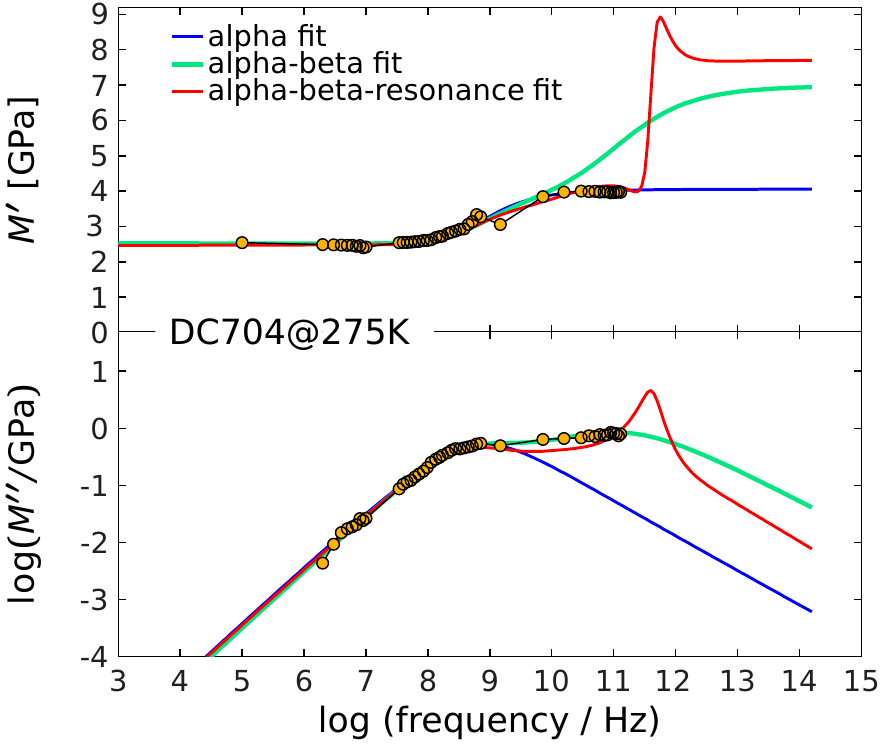}
  \caption{ \label{fig:kramerskronigcheck_dc704}The complex
    longitudinal modulus of DC704 at 275~K including several fitting
    schemes: the blue curve is a stretched exponential (Eq.\
    (\ref{eq:KWW})) fit of the real part, the green curve is an
    alpha-beta model fit \cite{sSaglanmak2010,sJakobsen2011a} to the
    imaginary part, and the red curve is a an alpha-beta model fit
    with an added resonance.}
\end{figure}
   
The KWW form describes only alpha relaxation, not the faster beta
relaxation dynamics which clearly appear in the imaginary compliance
spectra $J''(\omega)$ at high frequencies. In order to extend our fits
to this part of the spectra, we used an alpha-beta merging model that
has been described elsewhere\cite{sSaglanmak2010,sJakobsen2011a}. The
fitted alpha and beta relaxation times from this model are shown in in
Fig.~\ref{fig:alphabeta}. The alpha relaxation times show excellent
agreement with the $\tau_\alpha$ values obtained from the stretched
exponential fits. The beta relaxation time is nearly constant over the
range of temperatures explored here, i.e., $\tau_\beta\sim1$~ps, as
expected. Although the alpha-beta model fits were not used to test
MCT, they can provide insights based on their parameter values and on
their implications for the connections between the real and imaginary
parts of the compliance. Both the KWW and the alpha-beta fits are
illustrated in Fig.~\ref{fig:kramerskronigcheck_dc704} for a single
temperature, 275~K. The alpha-beta model fit overshoots the real part
dramatically at the highest frequencies. However, the resonance region
is approached here, and we can add this to our alpha-beta model and
obtain a curve that matches the rise in the imaginary part while the
real part remains flat. This is shown in Fig.\
\ref{fig:kramerskronigcheck_dc704} in red. The fit is not perfect; the
purpose here is merely to show that although beta relaxation is
apparent in the imaginary spectrum only, the data do not violate the
Kramers-Kronig relations. Owing to the consistency of the KWW fitting
scheme with the alpha-beta analysis, as shown in
Fig.~\ref{fig:alphabeta}, the KWW fits were used in Fig.\ 3(b) of the
main text and the subsequent analysis.

\subsection{Temperature dependent compliance minimum values}

\begin{figure}
 \includegraphics[height=5.6cm]{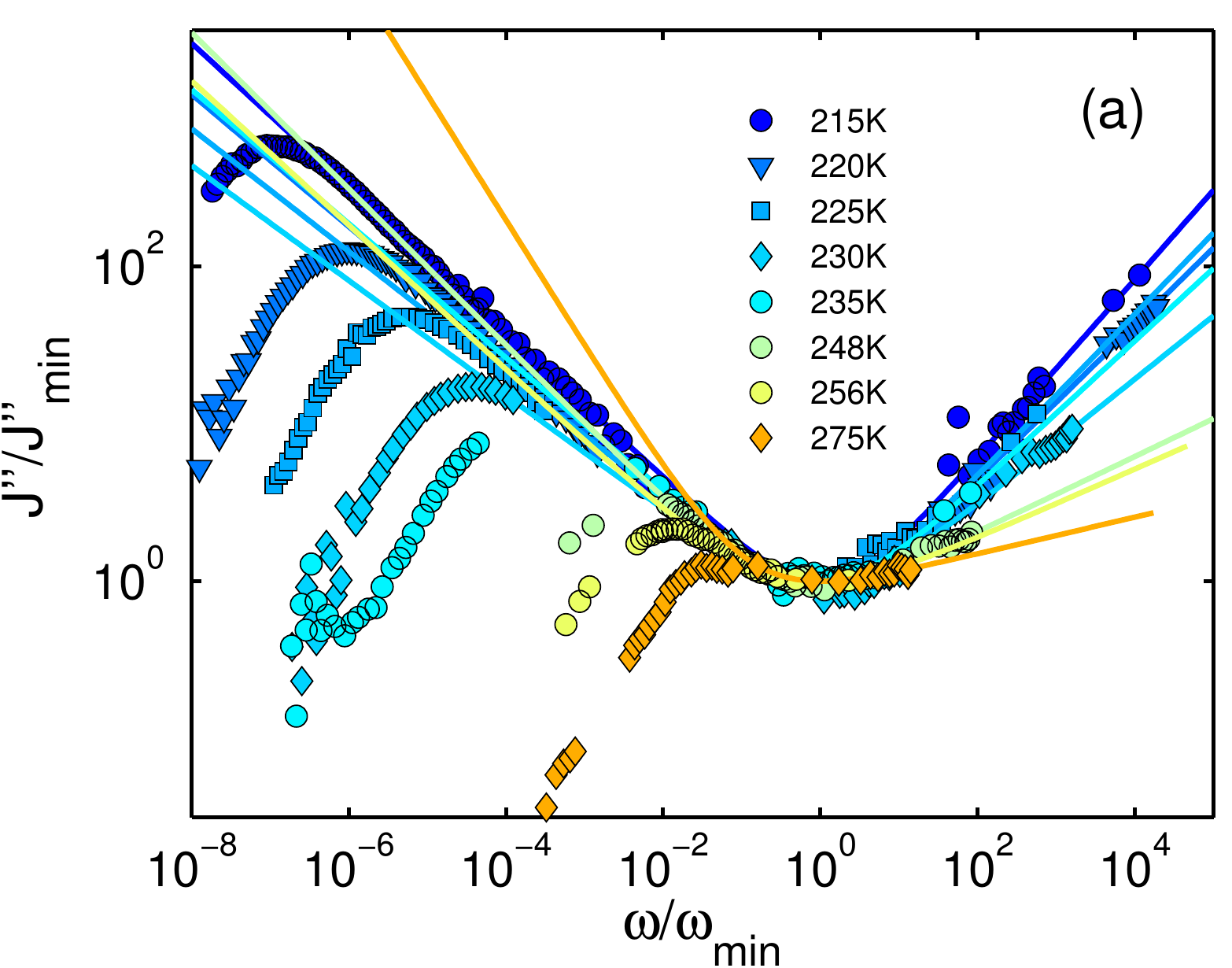}
 \includegraphics[height=5.8cm]{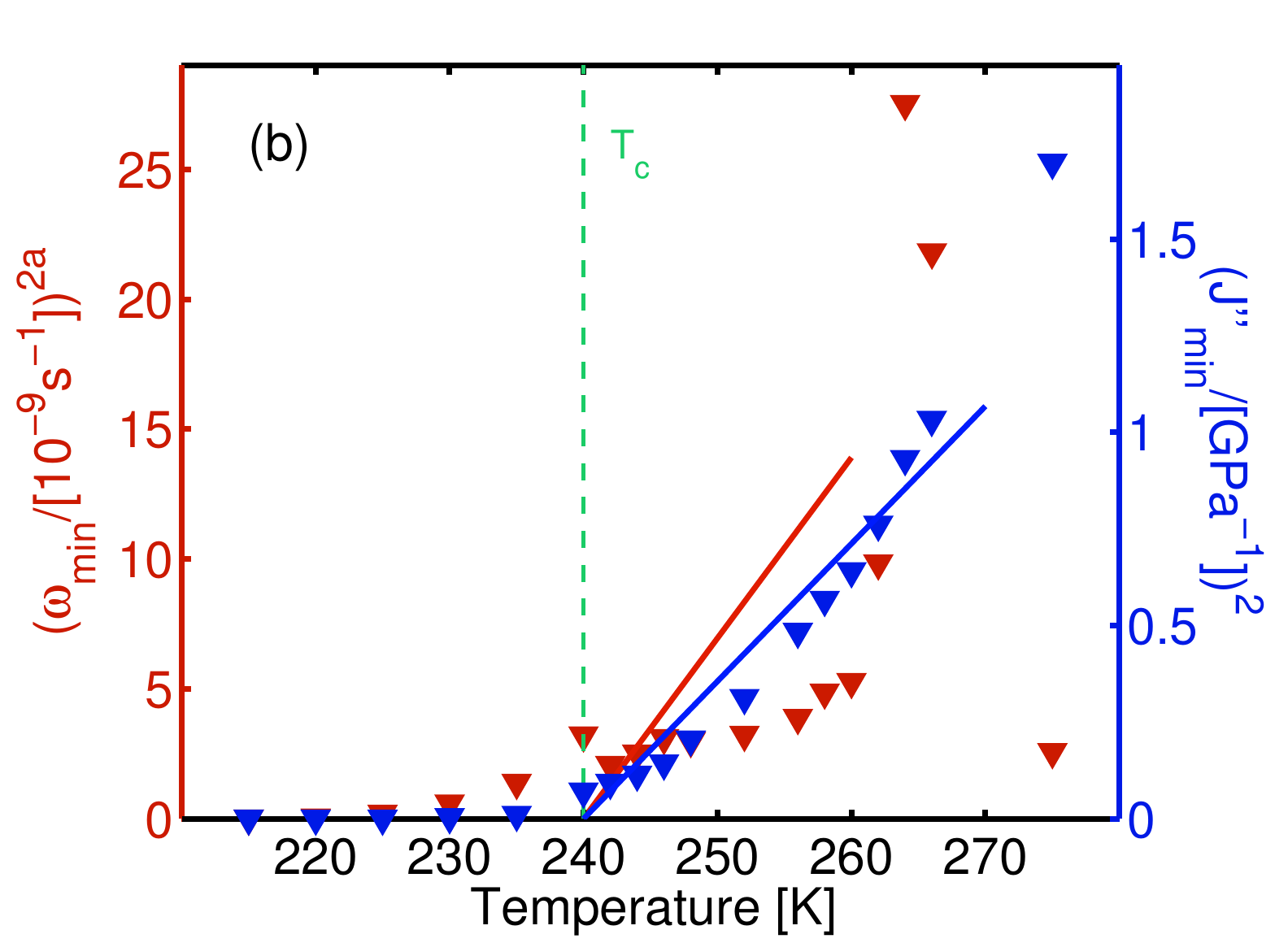}
 \caption{\label{mct_pl_pred}Test of mode-coupling theory predictions
   for the evolution of the susceptibility minimum
   $J^{\prime\prime}_{\text{min}}(\omega_{\text{min}})$. (a)
   $J''(\omega)$ rescaled at the susceptibility minimum. The fits are
   given to the asymptotic form given by Eq.\ (3) of the main text.
   (b) $(\omega_{\text{min}})^{2a}$ and
   $(J^{\prime\prime}_{\text{min}})^{2}$ vs $T$. $T_c=240$~K is marked
   by the vertical dashed line, and linear fits (keeping $T_c$ fixed)
   to temperatures in the range from 240~K to 248~K are shown as solid
   lines.}
\end{figure}

Using the fits described in the previous section, the imaginary
compliance $J''(\omega)$ plots may be superposed upon each other by
scaling the axes by the frequency minimum $\omega_{\text{min}}$ and
the imaginary compliance value $J''_{\text{min}}$, shown in Fig.\
\ref{mct_pl_pred}(a). The determined frequencies
$\omega_\text{min}(T)$ and compliance values $J''_\text{min}(T)$ at
the minima between the alpha and beta relaxation features may be used
in conjunction with the dynamic exponent $a$ to test additional MCT
predictions. These predictions indicate the scaling of the imaginary
susceptibility minimum as $\chi''(\omega_{\textrm{min}}) \propto
\left| T - T_c \right| ^{1/2}$ and of the minimum frequency as
$\omega_{\textrm{min}} \propto \left| T - T_c \right|^{1/2a}$. In
Fig.\ \ref{mct_pl_pred}(b) we plot both $\left( J''_{\textrm{min}}
\right)^2$ and $\left( \omega_{\textrm{min}}(T) \right)^{2a}$ versus
temperature. The lines shown are the predictions of MCT based on the
value $T_c=240$~K. As in Figs.~5(b)--5(d) of the main paper, the
parameters can be determined over a wide temperature range, but only
the 240--248~K range is useful for comparison to MCT
predictions. However, the uncertainties in $\omega_{\textrm{min}}$ and
$J''_{\textrm{min}}$ make this assessment difficult.

\clearpage
%


\begin{thebibliography}{10}

\bibitem{Kauzmann1948}
Kauzmann, W.
\newblock (1948) The nature of the glassy state and the behavior of liquids at
  low temperatures.
\newblock {\em Chemical Review} {\bf 43}, 219--256.

\bibitem{Angell1995}
Angell, C.~A.
\newblock (1995) Formation of glasses from liquids and biopolymers.
\newblock {\em Science} {\bf 267}, 1924--1935.

\bibitem{Debenedetti1996}
Debenedetti, P.~G.
\newblock (1996) {\em Metastable liquids : concepts and principles}.
\newblock (Princeton, N.J. : Princeton University Press).

\bibitem{Dyre2006}
Dyre, J.~C.
\newblock (2006) Solidity of viscous liquids. {IV}. density fluctuations.
\newblock {\em Phys. Rev. E} {\bf 74}, 021502.

\bibitem{Roland2010}
Roland, C.~M.
\newblock (2010) Relaxation phenomena in vitrifying polymers and molecular
  liquids.
\newblock {\em Macromolecules} {\bf 43}, 7875--7890.

\bibitem{Berthier2011}
Berthier, L \& Biroli, G.
\newblock (2011) Theoretical perspective on the glass transition and amorphous
  materials.
\newblock {\em Rev. Mod. Phys.} {\bf 83}, 587.

\bibitem{Johari1970}
Johari, G.~P \& Goldstein, M.
\newblock (1970) {Viscous Liquids and the Glass Transition. II. Secondary
  Relaxations in Glasses of Rigid Molecules}.
\newblock {\em J. Chem. Phys.} {\bf 53}, 2372.

\bibitem{Johari1982}
Johari, G.~P.
\newblock (1982) Effect of annealing on the secondary relaxations in glasses.
\newblock {\em J. Chem. Phys.} {\bf 77}, 4619--4626.

\bibitem{Dyre2006a}
Dyre, J.~C.
\newblock (2006) The glass transition and elastic models of glass-forming
  liquids.
\newblock {\em Reviews of Modern Physics} {\bf 78}, 953--972.

\bibitem{Gibbs1958}
Gibbs, J.~H \& DiMarzio, E.~A.
\newblock (1958) Nature of the glass transition and the glassy state.
\newblock {\em J. Chem. Phys.} {\bf 28}, 373--383.

\bibitem{Cohen1979}
Cohen, M.~H \& Grest, G.~S.
\newblock (1979) Liquid-glass transition, a free volume approach.
\newblock {\em Phys. Rev. B} {\bf 20}, 1077--1098.

\bibitem{Kivelson1995}
Kivelson, D, Kivelson, S.~A, Zhao, X.~L, Nussinov, Z,  \& Tarjus, G.
\newblock (1995) A thermodynamic theory of supercooled liquids.
\newblock {\em Physica A} {\bf 219}, 27--38.

\bibitem{Jung2005}
Jung, Y, Garrahan, J.~P,  \& Chandler, D.
\newblock (2005) Dynamical exchanges in facilitated models of supercooled
  liquids.
\newblock {\em J. Chem. Phys.} {\bf 123}.

\bibitem{Tripathy2009}
Tripathy, M \& Schweizer, K.~S.
\newblock (2009) The influence of shape on the glassy dynamics of hard
  nonspherical particle fluids. {II}. barriers, relaxation, fragility, kinetic
  vitrification, and universality.
\newblock {\em J. Chem. Phys.} {\bf 130}, 244907.

\bibitem{Gotze1992}
G\"{o}tze, W \& Sj\"{o}gren, L.
\newblock (1992) Relaxation processes in supercooled liquids.
\newblock {\em Rep. Prog. Phys.} {\bf 55}, 241--376.

\bibitem{Das2004}
Das, S.~P.
\newblock (2004) Mode-coupling theory and the glass transition in supercooled
  liquids.
\newblock {\em Rev. Mod. Phys.} {\bf 76}, 785--851.

\bibitem{Ferry1980}
Ferry, J.~D.
\newblock (1980) {\em Viscoelastic properties of polymers}.
\newblock (John Wiley \& Sons, Inc).

\bibitem{Plazek1996}
Plazek, D.~J.
\newblock (1996) 1995 {B}ingham {M}edal {A}ddress: Oh, thermorheological
  simplicity, wherefore art thou?
\newblock {\em Journal of Rheology} {\bf 40}, 987.

\bibitem{Olsen2001}
Olsen, N.~B, Christensen, T,  \& Dyre, J.~C.
\newblock (2001) Time-temperature superposition in viscous liquids.
\newblock {\em Phys. Rev. Lett.} {\bf 86}, 1271--1274.

\bibitem{Narayanaswamy1971}
Narayanaswamy, O.~S.
\newblock (1971) A model of structural relaxation in glass.
\newblock {\em J. Am. Ceram. Soc.} {\bf 54}, 491--498.

\bibitem{Moynihan1976}
Moynihan, C.~T, Easteal, A.~J, DeBolt, M,  \& Tucker, J.
\newblock (1976) Dependence of the fictive temperature of glass on cooling
  rate.
\newblock {\em American Ceramic Society} {\bf 59}, 12--16.

\bibitem{Lunkenheimer2000}
Lunkenheimer, P, Schneider, U, Brand, R,  \& Loidl, A.
\newblock (2000) Glassy dynamics.
\newblock {\em Contemp. Phys.} {\bf 41}, 15--36.

\bibitem{Petzold2010}
Petzold, N \& R\"{o}ssler, E.~A.
\newblock (2010) Light scattering study on the glass former o-terphenyl.
\newblock {\em J. Chem. Phys.} {\bf 133}, 124512.

\bibitem{Li1992}
Li, G, Du, W.~M, Sakai, A,  \& Cummins, H.~Z.
\newblock (1992) Light-scattering investigation of \ensuremath{\alpha} and
  \ensuremath{\beta} relaxation near the liquid-glass transition of the
  molecular glass salol.
\newblock {\em Phys. Rev. A} {\bf 46}, 3343--3356.

\bibitem{Li1992a}
Li, G, Du, W.~M, Chen, X.~K, Cummins, H.~Z,  \& Tao, N.~J.
\newblock (1992) Testing mode-coupling predictions for \ensuremath{\alpha} and
  \ensuremath{\beta} relaxation in ({C}a$_{0.4}${K}$_{0.6}$({NO}$_{3}$)$_{1.4}$
  near the liquid-glass transition by light scattering.
\newblock {\em Phys. Rev. A} {\bf 45}, 3867--3879.

\bibitem{Gotze1999}
G{\"o}tze, W.
\newblock (1999) Recent tests of the mode-coupling theory for glassy dynamics.
\newblock {\em J. Phys.: Condens. Matter} {\bf 11}, A1.

\bibitem{Wuttke1994}
Wuttke, J, Hernandez, J, Li, G, Coddens, G, Cummins, H.~Z, Fujiara, F, Petry,
  W,  \& Sillescu, H.
\newblock (1994) Neutron and light-scattering study of supercooled glycerol.
\newblock {\em Phys. Rev. Lett.} {\bf 72}, 3052--3055.

\bibitem{Shen2000}
Shen, G.~Q, Toulouse, J, Beaufils, S, Bonello, B, Hwang, Y.~H, Finkel, P,
  Hernandez, J, Bertault, M, Maglione, M, Ecolivet, C,  \& Cummins, H.~Z.
\newblock (2000) Experimental studies of the liquid-glass transition in
  trimethylheptane.
\newblock {\em Phys. Rev. E} {\bf 62}, 783--792.

\bibitem{Yan1987a}
Yan, Y.-X \& Nelson, K.~A.
\newblock (1987) Impulsive stimulated light scattering. {I}. {G}eneral theory.
\newblock {\em J. Chem. Phys.} {\bf 87}, 6240--6256.

\bibitem{Silence1992}
Silence, S.~M, Duggal, A.~R, Dhar, L,  \& Nelson, K.~A.
\newblock (1992) Structural and orientational relaxation in supercooled liquid
  triphenylphosphite.
\newblock {\em J. Chem. Phys.} {\bf 96}, 5448--5459.

\bibitem{Thomsen1986}
Thomsen, C, Grahn, H, Maris, H,  \& Tauc, J.
\newblock (1986) Surface generation and detection of phonons by picosecond
  light pulses.
\newblock {\em Phys. Rev. B} {\bf 34}, 4129--4138.

\bibitem{Choi2005}
Choi, J.~D, Feurer, T, Yamaguchi, M, Paxton, B,  \& Nelson, K.~A.
\newblock (2005) Generation of ultrahigh-frequency tunable acoustic waves.
\newblock {\em Appl. Phys. Lett.} {\bf 87}, 081907.

\bibitem{Supplementary}
(year?) See online supplementary information.

\bibitem{Hecksher2013}
Hecksher, T, Olsen, N.~B, Nelson, K.~A, Dyre, J.~C,  \& Christensen, T.
\newblock (2013) {Mechanical spectrum of viscous liquids. I. Low-frequency bulk
  and shear moduli of DC704 and 5-PPE measured by piezoceramic transducers}.
\newblock {\em J. Chem. Phys.} {\bf 138}, 12A543.

\bibitem{Christensen1994b}
Christensen, T \& Olsen, N.~B.
\newblock (1994) Determination of the frequency-dependent bulk modulus of
  glycerol using a piezoelectric spherical shell.
\newblock {\em Phys. Rev. B} {\bf 49}, 15396--15399.

\bibitem{Christensen1995}
Christensen, T \& Olsen, N.~B.
\newblock (1995) A rheometer for the measurement of high shear modulus covering
  more than seven decades of frequency below 50 khz.
\newblock {\em Rev. Sci. Instrum.} {\bf 66}, 5019--5031.

\bibitem{Klieber2013}
Klieber, C, Hecksher, T, Pezeril, T, Torchinsky, D.~H, Dyre, J.~C,  \& Nelson,
  K.~A.
\newblock (2013) Mechanical spectra of viscous liquids. {II}.
  {G}igahertz-frequency longitudinal and shear acoustic dynamics in glycerol
  and dc704 studied by time-domain brillouin scattering.
\newblock {\em J. Chem. Phys.} {\bf 138}, 12A544.

\bibitem{Barlow1967}
Barlow, A.~J, Erginsav, A,  \& Lamb, J.
\newblock (1967).
\newblock {\em Proc. R. Soc. London, Ser. A} {\bf 298}, 461.

\bibitem{Nielsen2009}
Nielsen, A.~I, Jakobsen, B, Niss, K, Olsen, N.~B, Richert, R,  \& Dyre, J.~C.
\newblock (2009) Prevalence of approximate $\sqrt{t}$ relaxation for the
  dielectric $\alpha$ process in visous organic liquids.
\newblock {\em J. Chem. Phys.} {\bf 130}, 154508.

\bibitem{Bohmer1993}
B\"{o}hmer, R, Ngai, K.~L, Angell, C.~A,  \& Plazek, D.~J.
\newblock (1993) Nonexponential relaxation in strong and fragile glass-formers.
\newblock {\em J. Chem. Phys.} {\bf 99}, 4201.

\bibitem{Vogel1921}
Vogel, H.
\newblock (1921) Das {T}emperaturabh\"{a}ngigkeitsgesetz der {V}iskosit\"{a}t
  von {F}l\"{u}ssigkeiten.
\newblock {\em Phys. Z.} {\bf 22}, 645--646.

\bibitem{Tammann1925}
Tammann, G.
\newblock (1925) Glasses as supercooled liquids.
\newblock {\em J. Soc. Glass Technol.} {\bf 9}, 166--185.

\bibitem{Cummins1993}
Cummins, H.~Z, Du, W.~M, Fuchs, M, G\"otze, W, A.Latz, Li, G,  \& Tao, N.~J.
\newblock (1993) Light scattering spectroscopy of the liquid-glass transition:
  comparison with idealized and extended mode coupling theory.
\newblock {\em Physica A: Statistical Mechanics and its Applications} {\bf
  201}, 207 -- 222.

\bibitem{Goldstein1969}
Goldstein, M.
\newblock (1969) Viscous liquids and the glass transition: {A} potential energy
  barrier picture.
\newblock {\em J. Chem. Phys.} {\bf 51}, 3728--3739.

\bibitem{Debenedetti2001}
Debenedetti, P.~G \& Stillinger, F.~H.
\newblock (2001) Supercooled liquids and the glass transition.
\newblock {\em Nature} {\bf 410}, 259--267.

\bibitem{Schroder2000}
Schroeder, T.~B, Sastry, S, Dyre, J,  \& Glotzer, S.~C.
\newblock (2000) Crossover to potential energy landscape dominated dynamics in
  a model glass-forming liquid.
\newblock {\em J. Chem. Phys.} {\bf 112}, 9834--9840.

\bibitem{Cummins1999}
Cummins, H.
\newblock (1999) The liquid-glass transition: a mode-coupling perspective.
\newblock {\em J. Phys.: Condens. Matter} {\bf 11}, A95.

\bibitem{Torchinsky2012}
Torchinsky, D.~H, Johnson, J.~A,  \& Nelson, K.~A.
\newblock (2012) $\alpha$-scale decoupling of the mechanical relaxation and
  diverging shear wave propagation length scale in triphenylphosphite.
\newblock {\em J. Chem. Phys.} {\bf 136}, 174509.

\bibitem{Hecksher_thesis}
Hecksher, T.
\newblock (2010) Ph.D. thesis (Roskilde University).

\bibitem{Johnson_thesis}
Johnson, J.~A.
\newblock (2011) Ph.D. thesis (Massachusetts Institute of Technology).

\bibitem{Yan1987b}
Yan, Y.-X \& Nelson, K.~A.
\newblock (1987) Impulsive stimulated light scattering. {II}. {C}omparison to
  frequency-domain light-scattering spectroscopy.
\newblock {\em J. Chem. Phys.} {\bf 87}, 6257--6266.

\bibitem{Torchinsky_thesis}
Torchinsky, D.~H.
\newblock (2008) {PhD} thesis (Massachusetts Institute of Technology,
  Cambridge, {MA}).

\bibitem{Klieber_thesis}
Klieber, C.
\newblock (2010) Ph.D. thesis (Massachusetts Institute of Technology).

\bibitem{Orcutt1973}
Orcutt, R.~H.
\newblock (1973) Interlot density variation of a siloxane manometer fluid.
\newblock {\em Journal of Vacuum Science Technology} {\bf 10}, 506--506.

\bibitem{Poulter1979}
Poulter, K.~F \& Nash, P.~J.
\newblock (1979) Interferometric oil micromanometer.
\newblock {\em Journal of Physics E-Scientific Instruments} {\bf 12}, 931--936.

\bibitem{Christensen1994}
Christensen, T \& Olsen, N.~B.
\newblock (1994) Quasistatic measurements of the frequency-dependent bulk and
  shear modulus of supercooled liquids.
\newblock {\em J. Non-Cryst. Solids} {\bf 172-174}, 362--364.

\bibitem{Gundermann2014}
Gundermann, D, Niss, K, Christensen, T, Dyre, J.~C,  \& Hecksher, T.
\newblock (2014) The dynamic bulk modulus of three glass-forming liquids.
\newblock {\em J. Chem. Phys.} {\bf 140}, 244508.

\bibitem{Igarashi2008a}
Igarashi, B, Christensen, T, Larsen, E.~H, Olsen, N.~B, Pedersen, I.~H,
  Rasmussen, T,  \& Dyre, J.~C.
\newblock (2008) A cryostat and temperature control system optimized for
  measuring relaxations of glass-forming liquids.
\newblock {\em Rev. Sci. Instrum.} {\bf 79}, 045105.

\end{thebibliography}

\begin{thebibliography}{27}%
\makeatletter
\providecommand \@ifxundefined [1]{%
 \@ifx{#1\undefined}
}%
\providecommand \@ifnum [1]{%
 \ifnum #1\expandafter \@firstoftwo
 \else \expandafter \@secondoftwo
 \fi
}%
\providecommand \@ifx [1]{%
 \ifx #1\expandafter \@firstoftwo
 \else \expandafter \@secondoftwo
 \fi
}%
\providecommand \natexlab [1]{#1}%
\providecommand \enquote  [1]{``#1''}%
\providecommand \bibnamefont  [1]{#1}%
\providecommand \bibfnamefont [1]{#1}%
\providecommand \citenamefont [1]{#1}%
\providecommand \href@noop [0]{\@secondoftwo}%
\providecommand \href [0]{\begingroup \@sanitize@url \@href}%
\providecommand \@href[1]{\@@startlink{#1}\@@href}%
\providecommand \@@href[1]{\endgroup#1\@@endlink}%
\providecommand \@sanitize@url [0]{\catcode `\\12\catcode `\$12\catcode
  `\&12\catcode `\#12\catcode `\^12\catcode `\_12\catcode `\%12\relax}%
\providecommand \@@startlink[1]{}%
\providecommand \@@endlink[0]{}%
\providecommand \url  [0]{\begingroup\@sanitize@url \@url }%
\providecommand \@url [1]{\endgroup\@href {#1}{\urlprefix }}%
\providecommand \urlprefix  [0]{URL }%
\providecommand \Eprint [0]{\href }%
\providecommand \doibase [0]{http://dx.doi.org/}%
\providecommand \selectlanguage [0]{\@gobble}%
\providecommand \bibinfo  [0]{\@secondoftwo}%
\providecommand \bibfield  [0]{\@secondoftwo}%
\providecommand \translation [1]{[#1]}%
\providecommand \BibitemOpen [0]{}%
\providecommand \bibitemStop [0]{}%
\providecommand \bibitemNoStop [0]{.\EOS\space}%
\providecommand \EOS [0]{\spacefactor3000\relax}%
\providecommand \BibitemShut  [1]{\csname bibitem#1\endcsname}%
\let\auto@bib@innerbib\@empty
\bibitem [{\citenamefont {Hecksher}\ \emph {et~al.}(2013)\citenamefont
  {Hecksher}, \citenamefont {Olsen}, \citenamefont {Nelson}, \citenamefont
  {Dyre},\ and\ \citenamefont {Christensen}}]{sHecksher2013}%
  \BibitemOpen
  \bibfield  {author} {\bibinfo {author} {\bibfnamefont {T.}~\bibnamefont
  {Hecksher}}, \bibinfo {author} {\bibfnamefont {N.~B.}\ \bibnamefont {Olsen}},
  \bibinfo {author} {\bibfnamefont {K.~A.}\ \bibnamefont {Nelson}}, \bibinfo
  {author} {\bibfnamefont {J.~C.}\ \bibnamefont {Dyre}}, \ and\ \bibinfo
  {author} {\bibfnamefont {T.}~\bibnamefont {Christensen}},\ }\href@noop {}
  {\bibfield  {journal} {\bibinfo  {journal} {J. Chem. Phys.}\ }\textbf
  {\bibinfo {volume} {138}},\ \bibinfo {pages} {12A543} (\bibinfo {year}
  {2013})}\BibitemShut {NoStop}%

\bibitem [{\citenamefont {Christensen}\ and\ \citenamefont
  {Olsen}(1994)}]{sChristensen1994b}%
  \BibitemOpen
  \bibfield  {author} {\bibinfo {author} {\bibfnamefont {T.}~\bibnamefont
  {Christensen}}\ and\ \bibinfo {author} {\bibfnamefont {N.~B.}\ \bibnamefont
  {Olsen}},\ }\href@noop {} {\bibfield  {journal} {\bibinfo  {journal} {Phys.
  Rev. B}\ }\textbf {\bibinfo {volume} {49}},\ \bibinfo {pages} {15396}
  (\bibinfo {year} {1994})}\BibitemShut {NoStop}%

\bibitem [{\citenamefont {Christensen}\ and\ \citenamefont
  {Olsen}(1995)}]{sChristensen1995}%
  \BibitemOpen
  \bibfield  {author} {\bibinfo {author} {\bibfnamefont {T.}~\bibnamefont
  {Christensen}}\ and\ \bibinfo {author} {\bibfnamefont {N.~B.}\ \bibnamefont
  {Olsen}},\ }\href@noop {} {\bibfield  {journal} {\bibinfo  {journal} {Rev.
  Sci. Instrum.}\ }\textbf {\bibinfo {volume} {66}},\ \bibinfo {pages} {5019}
  (\bibinfo {year} {1995})}\BibitemShut {NoStop}%

\bibitem [{\citenamefont {Hecksher}(2010)}]{sHecksher_thesis}%
  \BibitemOpen
  \bibfield  {author} {\bibinfo {author} {\bibfnamefont {T.}~\bibnamefont
  {Hecksher}},\ }\emph {\bibinfo {title} {Relaxation in supercooled
  liquids.}},\ \href@noop {} {Ph.D. thesis},\ \bibinfo  {school} {Roskilde
  University} (\bibinfo {year} {2010})\BibitemShut {NoStop}%

\bibitem [{\citenamefont {Johnson}(2011)}]{sJohnson_thesis}%
  \BibitemOpen
  \bibfield  {author} {\bibinfo {author} {\bibfnamefont {J.~A.}\ \bibnamefont
  {Johnson}},\ }\emph {\bibinfo {title} {Optical Characterization of Complex
  Mechanical and Thermal Transport Properties}},\ \href@noop {} {Ph.D.
  thesis},\ \bibinfo  {school} {Massachusetts Institute of Technology}
  (\bibinfo {year} {2011})\BibitemShut {NoStop}%

\bibitem [{\citenamefont {Yan}\ and\ \citenamefont
  {Nelson}(1987{\natexlab{a}})}]{sYan1987b}%
  \BibitemOpen
  \bibfield  {author} {\bibinfo {author} {\bibfnamefont {Y.-X.}\ \bibnamefont
  {Yan}}\ and\ \bibinfo {author} {\bibfnamefont {K.~A.}\ \bibnamefont
  {Nelson}},\ }\href@noop {} {\bibfield  {journal} {\bibinfo  {journal} {J.
  Chem. Phys.}\ }\textbf {\bibinfo {volume} {87}},\ \bibinfo {pages} {6257}
  (\bibinfo {year} {1987}{\natexlab{a}})}\BibitemShut {NoStop}%

\bibitem [{\citenamefont {Silence}\ \emph {et~al.}(1992)\citenamefont
  {Silence}, \citenamefont {Duggal}, \citenamefont {Dhar},\ and\ \citenamefont
  {Nelson}}]{sSilence1992}%
  \BibitemOpen
  \bibfield  {author} {\bibinfo {author} {\bibfnamefont {S.~M.}\ \bibnamefont
  {Silence}}, \bibinfo {author} {\bibfnamefont {A.~R.}\ \bibnamefont {Duggal}},
  \bibinfo {author} {\bibfnamefont {L.}~\bibnamefont {Dhar}}, \ and\ \bibinfo
  {author} {\bibfnamefont {K.~A.}\ \bibnamefont {Nelson}},\ }\href@noop {}
  {\bibfield  {journal} {\bibinfo  {journal} {J. Chem. Phys.}\ }\textbf
  {\bibinfo {volume} {96}},\ \bibinfo {pages} {5448} (\bibinfo {year}
  {1992})}\BibitemShut {NoStop}%

\bibitem [{\citenamefont {Torchinsky}(2008)}]{sTorchinsky_thesis}%
  \BibitemOpen
  \bibfield  {author} {\bibinfo {author} {\bibfnamefont {D.~H.}\ \bibnamefont
  {Torchinsky}},\ }\emph {\bibinfo {title} {Optical Study of Shear and
  Longitudinal Acoustic Waves and Complex Relaxation Dynamics of Glass Forming
  Liquids}},\ \href@noop {} {\bibinfo {type} {{PhD} thesis}},\ \bibinfo
  {school} {Massachusetts Institute of Technology}, \bibinfo {address}
  {Cambridge, {MA}} (\bibinfo {year} {2008})\BibitemShut {NoStop}%

\bibitem [{\citenamefont {Thomsen}\ \emph {et~al.}(1986)\citenamefont
  {Thomsen}, \citenamefont {Grahn}, \citenamefont {Maris},\ and\ \citenamefont
  {Tauc}}]{sThomsen1986}%
  \BibitemOpen
  \bibfield  {author} {\bibinfo {author} {\bibfnamefont {C.}~\bibnamefont
  {Thomsen}}, \bibinfo {author} {\bibfnamefont {H.}~\bibnamefont {Grahn}},
  \bibinfo {author} {\bibfnamefont {H.}~\bibnamefont {Maris}}, \ and\ \bibinfo
  {author} {\bibfnamefont {J.}~\bibnamefont {Tauc}},\ }\href@noop {} {\bibfield
   {journal} {\bibinfo  {journal} {Phys. Rev. B}\ }\textbf {\bibinfo {volume}
  {34}},\ \bibinfo {pages} {4129} (\bibinfo {year} {1986})}\BibitemShut
  {NoStop}%

\bibitem [{\citenamefont {Klieber}(2010)}]{sKlieber_thesis}%
  \BibitemOpen
  \bibfield  {author} {\bibinfo {author} {\bibfnamefont {C.}~\bibnamefont
  {Klieber}},\ }\emph {\bibinfo {title} {Ultrafast photo-acoustic spectroscopy
  of supercooled liquids}},\ \href@noop {} {Ph.D. thesis},\ \bibinfo  {school}
  {Massachusetts Institute of Technology} (\bibinfo {year} {2010})\BibitemShut
  {NoStop}%

\bibitem [{\citenamefont {Choi}\ \emph {et~al.}(2005)\citenamefont {Choi},
  \citenamefont {Feurer}, \citenamefont {Yamaguchi}, \citenamefont {Paxton},\
  and\ \citenamefont {Nelson}}]{sChoi2005}%
  \BibitemOpen
  \bibfield  {author} {\bibinfo {author} {\bibfnamefont {J.~D.}\ \bibnamefont
  {Choi}}, \bibinfo {author} {\bibfnamefont {T.}~\bibnamefont {Feurer}},
  \bibinfo {author} {\bibfnamefont {M.}~\bibnamefont {Yamaguchi}}, \bibinfo
  {author} {\bibfnamefont {B.}~\bibnamefont {Paxton}}, \ and\ \bibinfo {author}
  {\bibfnamefont {K.~A.}\ \bibnamefont {Nelson}},\ }\href@noop {} {\bibfield
  {journal} {\bibinfo  {journal} {Appl. Phys. Lett.}\ }\textbf {\bibinfo
  {volume} {87}},\ \bibinfo {pages} {081907} (\bibinfo {year}
  {2005})}\BibitemShut {NoStop}%

\bibitem [{\citenamefont {Patel}\ and\ \citenamefont {Tam}(1981)}]{sPatel1981}%
  \BibitemOpen
  \bibfield  {author} {\bibinfo {author} {\bibfnamefont {C.~K.~N.}\
  \bibnamefont {Patel}}\ and\ \bibinfo {author} {\bibfnamefont {A.~C.}\
  \bibnamefont {Tam}},\ }\href@noop {} {\bibfield  {journal} {\bibinfo
  {journal} {Rev. Mod. Phys}\ }\textbf {\bibinfo {volume} {53}},\ \bibinfo
  {pages} {517} (\bibinfo {year} {1981})}\BibitemShut {NoStop}%

\bibitem [{\citenamefont {Neubrand}\ and\ \citenamefont
  {Hess}(1992)}]{sNeubrand1992}%
  \BibitemOpen
  \bibfield  {author} {\bibinfo {author} {\bibfnamefont {A.}~\bibnamefont
  {Neubrand}}\ and\ \bibinfo {author} {\bibfnamefont {P.}~\bibnamefont
  {Hess}},\ }\href@noop {} {\bibfield  {journal} {\bibinfo  {journal} {J. Appl.
  Phys.}\ }\textbf {\bibinfo {volume} {71}},\ \bibinfo {pages} {227} (\bibinfo
  {year} {1992})}\BibitemShut {NoStop}%

\bibitem [{\citenamefont {Hess}(1996)}]{sHess1996}%
  \BibitemOpen
  \bibfield  {author} {\bibinfo {author} {\bibfnamefont {P.}~\bibnamefont
  {Hess}},\ }\href@noop {} {\bibfield  {journal} {\bibinfo  {journal} {App.
  Surf. Sci.}\ }\textbf {\bibinfo {volume} {106}},\ \bibinfo {pages} {429}
  (\bibinfo {year} {1996})}\BibitemShut {NoStop}%

\bibitem [{\citenamefont {Glorieux}\ \emph {et~al.}(2004)\citenamefont
  {Glorieux}, \citenamefont {Beers}, \citenamefont {Bentefour}, \citenamefont
  {de~Rostyne},\ and\ \citenamefont {Nelson}}]{sGlorieux2004}%
  \BibitemOpen
  \bibfield  {author} {\bibinfo {author} {\bibfnamefont {C.}~\bibnamefont
  {Glorieux}}, \bibinfo {author} {\bibfnamefont {J.~D.}\ \bibnamefont {Beers}},
  \bibinfo {author} {\bibfnamefont {E.~H.}\ \bibnamefont {Bentefour}}, \bibinfo
  {author} {\bibfnamefont {K.~V.}\ \bibnamefont {de~Rostyne}}, \ and\ \bibinfo
  {author} {\bibfnamefont {K.~A.}\ \bibnamefont {Nelson}},\ }\href@noop {}
  {\bibfield  {journal} {\bibinfo  {journal} {Rev. Sci. Instr.}\ ,\ \bibinfo
  {pages} {2906}} (\bibinfo {year} {2004})}\BibitemShut {NoStop}%

\bibitem [{\citenamefont {Yang}\ and\ \citenamefont {Nelson}(1995)}]{sYang1995}%
  \BibitemOpen
  \bibfield  {author} {\bibinfo {author} {\bibfnamefont {Y.}~\bibnamefont
  {Yang}}\ and\ \bibinfo {author} {\bibfnamefont {K.~A.}\ \bibnamefont
  {Nelson}},\ }\href@noop {} {\bibfield  {journal} {\bibinfo  {journal} {J.
  Chem. Phys.}\ }\textbf {\bibinfo {volume} {103}},\ \bibinfo {pages} {7722}
  (\bibinfo {year} {1995})}\BibitemShut {NoStop}%

\bibitem [{\citenamefont {Hinze}\ \emph {et~al.}(2000)\citenamefont {Hinze},
  \citenamefont {Brace}, \citenamefont {Gottke},\ and\ \citenamefont
  {Fayer}}]{sHinze2000}%
  \BibitemOpen
  \bibfield  {author} {\bibinfo {author} {\bibfnamefont {G.}~\bibnamefont
  {Hinze}}, \bibinfo {author} {\bibfnamefont {D.~D.}\ \bibnamefont {Brace}},
  \bibinfo {author} {\bibfnamefont {S.~D.}\ \bibnamefont {Gottke}}, \ and\
  \bibinfo {author} {\bibfnamefont {M.~D.}\ \bibnamefont {Fayer}},\ }\href@noop
  {} {\bibfield  {journal} {\bibinfo  {journal} {J. Chem. Phys.}\ }\textbf
  {\bibinfo {volume} {113}},\ \bibinfo {pages} {3723} (\bibinfo {year}
  {2000})}\BibitemShut {NoStop}%

\bibitem [{\citenamefont {Yan}\ and\ \citenamefont
  {Nelson}(1987{\natexlab{b}})}]{sYan1987a}%
  \BibitemOpen
  \bibfield  {author} {\bibinfo {author} {\bibfnamefont {Y.-X.}\ \bibnamefont
  {Yan}}\ and\ \bibinfo {author} {\bibfnamefont {K.~A.}\ \bibnamefont
  {Nelson}},\ }\href@noop {} {\bibfield  {journal} {\bibinfo  {journal} {J.
  Chem. Phys.}\ }\textbf {\bibinfo {volume} {87}},\ \bibinfo {pages} {6240}
  (\bibinfo {year} {1987}{\natexlab{b}})}\BibitemShut {NoStop}%

\bibitem [{\citenamefont {Silence}(1991)}]{ssilence_thesis}%
  \BibitemOpen
  \bibfield  {author} {\bibinfo {author} {\bibfnamefont {S.}~\bibnamefont
  {Silence}},\ }\href@noop {} {Ph.D. thesis},\ \bibinfo  {school}
  {Massachusetts Institute of Technology} (\bibinfo {year} {1991})\BibitemShut
  {NoStop}%

\bibitem [{\citenamefont {Lin}\ \emph {et~al.}(1991)\citenamefont {Lin},
  \citenamefont {Stoner}, \citenamefont {Maris},\ and\ \citenamefont
  {Tauc}}]{sLin1991}%
  \BibitemOpen
  \bibfield  {author} {\bibinfo {author} {\bibfnamefont {H.~N.}\ \bibnamefont
  {Lin}}, \bibinfo {author} {\bibfnamefont {R.~J.}\ \bibnamefont {Stoner}},
  \bibinfo {author} {\bibfnamefont {H.~J.}\ \bibnamefont {Maris}}, \ and\
  \bibinfo {author} {\bibfnamefont {J.}~\bibnamefont {Tauc}},\ }\href@noop {}
  {\bibfield  {journal} {\bibinfo  {journal} {J. Appl. Phys.}\ }\textbf
  {\bibinfo {volume} {69}},\ \bibinfo {pages} {3816} (\bibinfo {year}
  {1991})}\BibitemShut {NoStop}%

\bibitem [{\citenamefont {Hurley}\ and\ \citenamefont
  {Wright}(1999)}]{sHurley1999}%
  \BibitemOpen
  \bibfield  {author} {\bibinfo {author} {\bibfnamefont {D.~H.}\ \bibnamefont
  {Hurley}}\ and\ \bibinfo {author} {\bibfnamefont {O.~B.}\ \bibnamefont
  {Wright}},\ }\href@noop {} {\bibfield  {journal} {\bibinfo  {journal} {Opt.
  Lett.}\ }\textbf {\bibinfo {volume} {24}},\ \bibinfo {pages} {1305(3)}
  (\bibinfo {year} {1999})}\BibitemShut {NoStop}%

\bibitem [{\citenamefont {Perrin}\ \emph {et~al.}(1999)\citenamefont {Perrin},
  \citenamefont {Rossignol}, \citenamefont {Bonello},\ and\ \citenamefont
  {Jeannet}}]{sPerrin1999}%
  \BibitemOpen
  \bibfield  {author} {\bibinfo {author} {\bibfnamefont {B.}~\bibnamefont
  {Perrin}}, \bibinfo {author} {\bibfnamefont {C.}~\bibnamefont {Rossignol}},
  \bibinfo {author} {\bibfnamefont {B.}~\bibnamefont {Bonello}}, \ and\
  \bibinfo {author} {\bibfnamefont {J.}~\bibnamefont {Jeannet}},\ }\href@noop
  {} {\bibfield  {journal} {\bibinfo  {journal} {Physica B: Cond. Matter}\
  }\textbf {\bibinfo {volume} {263-264}},\ \bibinfo {pages} {571} (\bibinfo
  {year} {1999})}\BibitemShut {NoStop}%

\bibitem [{\citenamefont {Wuttke}(2012)}]{sWuttke2012}%
  \BibitemOpen
  \bibfield  {author} {\bibinfo {author} {\bibfnamefont {J.}~\bibnamefont
  {Wuttke}},\ }\href@noop {} {\bibfield  {journal} {\bibinfo  {journal}
  {Algorithms}\ }\textbf {\bibinfo {volume} {5}},\ \bibinfo {pages} {604}
  (\bibinfo {year} {2012})}\BibitemShut {NoStop}%

\bibitem [{\citenamefont {Klieber}\ \emph {et~al.}(2013)\citenamefont
  {Klieber}, \citenamefont {Hecksher}, \citenamefont {Pezeril}, \citenamefont
  {Torchinsky}, \citenamefont {Dyre},\ and\ \citenamefont
  {Nelson}}]{sKlieber2013}%
  \BibitemOpen
  \bibfield  {author} {\bibinfo {author} {\bibfnamefont {C.}~\bibnamefont
  {Klieber}}, \bibinfo {author} {\bibfnamefont {T.}~\bibnamefont {Hecksher}},
  \bibinfo {author} {\bibfnamefont {T.}~\bibnamefont {Pezeril}}, \bibinfo
  {author} {\bibfnamefont {D.~H.}\ \bibnamefont {Torchinsky}}, \bibinfo
  {author} {\bibfnamefont {J.~C.}\ \bibnamefont {Dyre}}, \ and\ \bibinfo
  {author} {\bibfnamefont {K.~A.}\ \bibnamefont {Nelson}},\ }\href@noop {}
  {\bibfield  {journal} {\bibinfo  {journal} {J. Chem. Phys.}\ }\textbf
  {\bibinfo {volume} {138}},\ \bibinfo {pages} {12A544} (\bibinfo {year}
  {2013})}\BibitemShut {NoStop}%

\bibitem [{\citenamefont {Saglanmak}\ \emph {et~al.}(2010)\citenamefont
  {Saglanmak}, \citenamefont {Nielsen}, \citenamefont {Olsen}, \citenamefont
  {Dyre},\ and\ \citenamefont {Niss}}]{sSaglanmak2010}%
  \BibitemOpen
  \bibfield  {author} {\bibinfo {author} {\bibfnamefont {N.}~\bibnamefont
  {Saglanmak}}, \bibinfo {author} {\bibfnamefont {A.~I.}\ \bibnamefont
  {Nielsen}}, \bibinfo {author} {\bibfnamefont {N.~B.}\ \bibnamefont {Olsen}},
  \bibinfo {author} {\bibfnamefont {J.~C.}\ \bibnamefont {Dyre}}, \ and\
  \bibinfo {author} {\bibfnamefont {K.}~\bibnamefont {Niss}},\ }\href@noop {}
  {\bibfield  {journal} {\bibinfo  {journal} {J. Chem. Phys.}\ }\textbf
  {\bibinfo {volume} {132}},\ \bibinfo {pages} {024503} (\bibinfo {year}
  {2010})}\BibitemShut {NoStop}%

\bibitem [{\citenamefont {Jakobsen}\ \emph {et~al.}(2011)\citenamefont
  {Jakobsen}, \citenamefont {Niss}, \citenamefont {Maggi}, \citenamefont
  {Olsen}, \citenamefont {Christensen},\ and\ \citenamefont
  {Dyre}}]{sJakobsen2011a}%
  \BibitemOpen
  \bibfield  {author} {\bibinfo {author} {\bibfnamefont {B.}~\bibnamefont
  {Jakobsen}}, \bibinfo {author} {\bibfnamefont {K.}~\bibnamefont {Niss}},
  \bibinfo {author} {\bibfnamefont {C.}~\bibnamefont {Maggi}}, \bibinfo
  {author} {\bibfnamefont {N.~B.}\ \bibnamefont {Olsen}}, \bibinfo {author}
  {\bibfnamefont {T.}~\bibnamefont {Christensen}}, \ and\ \bibinfo {author}
  {\bibfnamefont {J.~C.}\ \bibnamefont {Dyre}},\ }\href@noop {} {\bibfield
  {journal} {\bibinfo  {journal} {J. Non-Cryst. Solids}\ }\textbf {\bibinfo
  {volume} {357}},\ \bibinfo {pages} {267} (\bibinfo {year}
  {2011})}\BibitemShut {NoStop}%

\bibitem [{\citenamefont {Jakobsen}\ \emph {et~al.}(2012)\citenamefont
  {Jakobsen}, \citenamefont {Hecksher}, \citenamefont {Christensen},
  \citenamefont {Olsen}, \citenamefont {Dyre},\ and\ \citenamefont
  {Niss}}]{sJakobsen2012}%
  \BibitemOpen
  \bibfield  {author} {\bibinfo {author} {\bibfnamefont {B.}~\bibnamefont
  {Jakobsen}}, \bibinfo {author} {\bibfnamefont {T.}~\bibnamefont {Hecksher}},
  \bibinfo {author} {\bibfnamefont {T.}~\bibnamefont {Christensen}}, \bibinfo
  {author} {\bibfnamefont {N.~B.}\ \bibnamefont {Olsen}}, \bibinfo {author}
  {\bibfnamefont {J.~C.}\ \bibnamefont {Dyre}}, \ and\ \bibinfo {author}
  {\bibfnamefont {K.}~\bibnamefont {Niss}},\ }\href@noop {} {\bibfield
  {journal} {\bibinfo  {journal} {J. Chem. Phys.}\ }\textbf {\bibinfo {volume}
  {136}},\ \bibinfo {pages} {081102} (\bibinfo {year} {2012})}\BibitemShut
  {NoStop}%
\end{thebibliography}

\end{document}